\documentclass[twocolumn,aps]{revtex4-1}
\usepackage{graphicx}
\usepackage{epsf}
\usepackage{amsmath}
\usepackage{amsfonts}
\usepackage{amssymb}

\def\vare{\varepsilon}
\def\av#1{\langle #1 \rangle}
\def\gsim{\lower -0.3ex \hbox{$>$} \kern -0.75em \lower 0.7ex
\hbox{$\sim$}}
\def\lsim{\lower -0.3ex \hbox{$<$} \kern -0.75em \lower 0.7ex \hbox{$\sim$}}
\def\Journal #1,#2,#3,#4#5#6#7{#1 {\bf #2}, #3 (#4#5#6#7)}

\begin{document}

\title{The electronic properties of bilayer graphene}
\author{Edward McCann}
\address{Department of Physics, Lancaster University, Lancaster, LA1 4YB, UK}

\author{Mikito Koshino}
\address{Department of Physics, Tohoku University, Sendai, 980-8578, Japan}

\begin{abstract}
We review the electronic properties of bilayer graphene,
beginning with a description of the tight-binding model of bilayer
graphene and the derivation of the effective Hamiltonian
describing massive chiral quasiparticles in two parabolic bands
at low energy. We take into account five tight-binding parameters of the
Slonczewski-Weiss-McClure model of bulk graphite plus intra- and
interlayer asymmetry between atomic sites which induce band gaps
in the low-energy spectrum. The Hartree model of screening and
band-gap opening due to interlayer asymmetry in the presence of
external gates is presented. The tight-binding model is used to
describe optical and transport properties including the integer
quantum Hall effect, and we also discuss orbital magnetism,
phonons and the influence of strain on electronic properties.
We conclude with an overview of electronic interaction effects.
\end{abstract}

\maketitle

\tableofcontents

\section{Introduction}

The production by mechanical exfoliation of isolated flakes of graphene with
excellent conducting properties \cite{novo04} was soon followed by the observation
of an unusual sequence of plateaus in the integer quantum Hall effect in monolayer graphene
\cite{novo05,zhang05}. This confirmed the fact that charge carriers in monolayer
graphene are massless chiral quasiparticles with a linear dispersion,
as described by a Dirac-like effective Hamiltonian \cite{d+m84,semenoff84,haldane88},
and it prompted an explosion of interest in the field \cite{cnreview}.

The integer quantum Hall effect in bilayer graphene \cite{novo06} is arguably even more
unusual than in monolayer because it indicates the presence of massive chiral quasiparticles \cite{mcc06a}
with a parabolic dispersion at low energy. The effective Hamiltonian of bilayer graphene
may be viewed as a generalisation of the Dirac-like Hamiltonian of monolayer graphene
and the second (after the monolayer) in a family of chiral Hamiltonians that appear
at low energy in ABC-stacked (rhombohedral) multilayer
graphene \cite{mcc06a,guinea06,koshino07,manes07,nak08,min08a,min08b}.
In addition to interesting underlying physics, bilayer graphene holds potential
for electronics applications, not least because of the possibility to control both
carrier density and energy band gap through doping or
gating \cite{mcc06a,guinea06,ohta06,mcc06b,min07,oostinga,castro}.


Not surprisingly, many of the properties of bilayer graphene are similar to those in
monolayer \cite{cnreview,novo12}. These include excellent electrical conductivity with room temperature
mobility of up to $40,000\,$cm$^{2}\,$V$^{-1}\,$s$^{-1}$ in air \cite{dean10};
the possibility to tune electrical properties by changing the carrier density through
gating or doping \cite{novo04,novo06,ohta06};
high thermal conductivity with room temperature thermal conductivity of about
$2,800\,$W$\,$m$^{-1}\,$K$^{-1}$ \cite{ghosh10,balandin11};
mechanical stiffness, strength and flexibility
(Young's modulus is estimated to be about $0.8\,$TPa \cite{neekamal10,zhang-wang11});
transparency with transmittance of white light of about $95\,\%$ \cite{nair08};
impermeability to gases \cite{bunch08};
and the ability to be chemically functionalised \cite{elias09}.
Thus, as with monolayer graphene, bilayer graphene has potential for future
applications in many areas \cite{novo12} including
transparent, flexible electrodes for touch screen displays \cite{bae10};
high-frequency transistors \cite{xia10};
thermoelectric devices \cite{wang11};
photonic devices including plasmonic devices \cite{yan12} and photodetectors \cite{yan-kim12};
energy applications including batteries \cite{sugawara11,kanetani12};
and composite materials \cite{gong12,young12}.

It should be stressed, however, that bilayer graphene has features
that make it distinct from monolayer. The low-energy band structure,
described in detail in Section~\ref{s:ebs}, is different.
Like monolayer, intrinsic bilayer has no band gap between its conduction and valence
bands, but the low-energy dispersion is quadratic (rather than linear as in monolayer)
with massive chiral quasiparticles \cite{novo06,mcc06a} rather than massless
ones.
As there are two layers, bilayer graphene represents the thinnest possible limit
of an intercalated material \cite{sugawara11,kanetani12}.
It is possible to address each layer separately leading
to entirely new functionalities in bilayer graphene including the possibility to control
an energy band gap of up to about $300\,$meV through doping or
gating \cite{mcc06a,guinea06,ohta06,mcc06b,min07,oostinga,castro}.
Recently, this band gap has been used to create devices - constrictions and dots - by
electrostatic confinement with gates \cite{goossens12}.
Bilayer or multilayer graphene devices may also be preferable to monolayer ones when there
is a need to use more material for increased electrical or thermal conduction,
strength \cite{gong12,young12}, or optical signature \cite{yan12}.


In the following we review the electronic properties of bilayer graphene.
Section~\ref{s:ebs} is an overview of the electronic tight-binding Hamiltonian and
resulting band structure describing the low-energy chiral Hamiltonian and taking into account
different parameters that couple atomic orbitals as well as external factors that may
change the electron bands by, for example, opening a band gap.
We include the Landau level spectrum in the presence of a perpendicular magnetic field
and the corresponding integer quantum Hall effect.
In section~\ref{s:tbg} we consider the opening of a band gap due to doping or gating
and present a simple analytical model that describes the density-dependence
of the band gap by taking into account screening by electrons on the bilayer device.
The tight-binding model is used to describe transport properties, section~\ref{s:tp},
and optical properties, section~\ref{sec_bi_optical}.
We also discuss orbital magnetism in section~\ref{s:om},
phonons and the influence of strain in section~\ref{s:pas}.
Section~\ref{s:eei} concludes with an overview of electronic-interaction effects.
Note that this review considers Bernal-stacked (also known as AB-stacked)
bilayer graphene; we do not consider other stacking types such as
AA-stacked graphene \cite{liu09}, twisted
graphene \cite{lopes07,berger06,hass08,mele10,li10,luican11}
or two graphene sheets separated by a dielectric with, possibly, electronic interactions
between them \cite{schmidt08,ni08,min08c,khar08,schmidt10,pon11}.


\section{Electronic band structure}\label{s:ebs}

\subsection{The crystal structure and the Brillouin zone}

\begin{figure}[t]
\centerline{\epsfxsize=0.7\hsize \epsffile{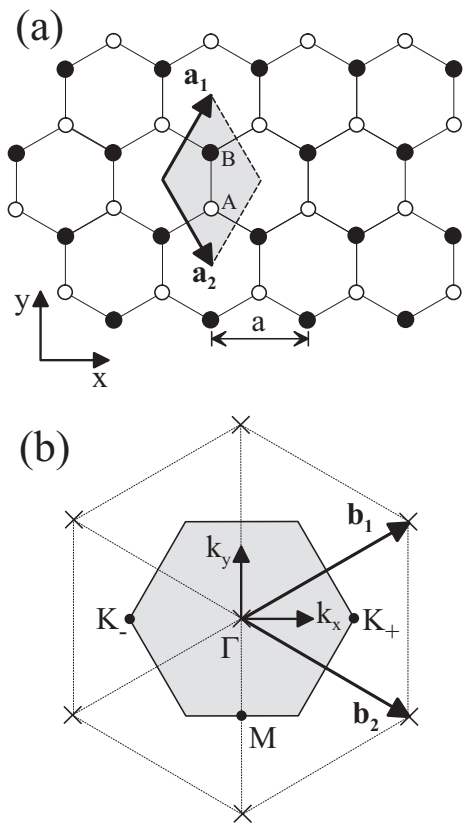}}
\caption{(a) Crystal structure of monolayer graphene with $A$ ($B$) atoms shown
as white (black) circles. The shaded rhombus is the conventional unit cell,
$\mathbf{a}_1$ and $\mathbf{a}_2$ are primitive lattice vectors.
(b) Reciprocal lattice of monolayer and bilayer graphene with lattice points
indicated as crosses, $\mathbf{b}_1$ and $\mathbf{b}_2$ are primitive reciprocal
lattice vectors. The shaded hexagon is the first Brillouin zone with $\Gamma$ indicating
the centre, and $K_{+}$ and $K_{-}$ showing two non-equivalent corners.}
\label{fig:monolattice}
\end{figure}

\begin{figure}[t]
\centerline{\epsfxsize=0.7\hsize \epsffile{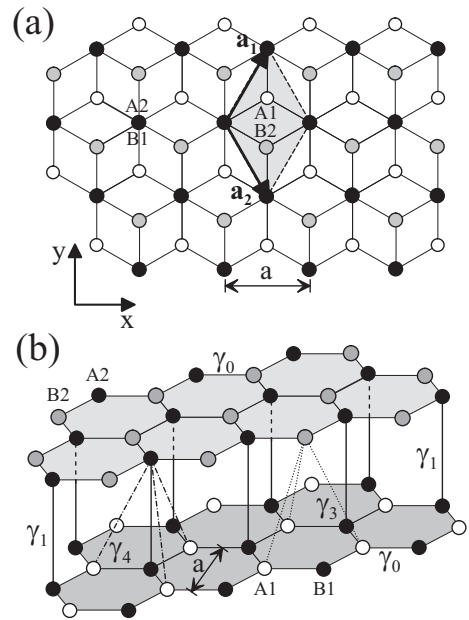}}
\caption{(a) Plan and (b) side view of the crystal structure of bilayer graphene.
Atoms $A1$ and $B1$ on the lower layer are shown as white and black circles,
$A2$, $B2$ on the upper layer are black and grey, respectively.
The shaded rhombus in (a) indicates the conventional unit cell.
}
\label{fig:bilayerlattice}
\end{figure}

Bilayer graphene consists of two coupled monolayers of carbon atoms, each
with a honeycomb crystal structure. Figure~\ref{fig:monolattice}
shows the crystal structure of monolayer graphene, figure~\ref{fig:bilayerlattice}
shows bilayer graphene.
In both cases, primitive lattice vectors
$\mathbf{a}_1$ and $\mathbf{a}_2$ may be defined as
\begin{eqnarray}
\mathbf{a}_1 = \left( \frac{a}{2} , \frac{\sqrt{3}a}{2} \right) \, , \qquad
\mathbf{a}_2 = \left( \frac{a}{2} , - \frac{\sqrt{3}a}{2} \right) \, ,
\end{eqnarray}
where $a = | \mathbf{a}_1 | = | \mathbf{a}_2 |$ is the lattice constant,
the distance between adjacent unit cells, $a = 2.46\,$\AA$\,$\cite{saito}.
Note that the lattice constant is distinct from the carbon-carbon bond length
$a_{CC} = a / \sqrt{3} = 1.42\,$\AA,  which is the distance between adjacent carbon atoms.

In monolayer graphene, each unit cell contains two carbon atoms, labelled
$A$ and $B$, figure~\ref{fig:monolattice}(a).
The positions of $A$ and $B$ atoms are not equivalent because it is not possible
to connect them with a lattice vector of the form
$\mathbf{R} = n_1 \mathbf{a}_1 + n_2 \mathbf{a}_2$, where $n_1$ and $n_2$ are integers.
Bilayer graphene consists of two coupled monolayers, with four atoms in the unit
cell, labelled $A1$, $B1$ on the lower layer and $A2$, $B2$ on the upper layer.
The layers are arranged so that one of the atoms from the lower layer $B1$ is directly
below an atom, $A2$, from the upper layer. We refer to these two atomic sites as `dimer' sites
because the electronic orbitals on them are coupled together by a relatively strong
interlayer coupling.
The other two atoms, $A1$ and $B2$, don't have
a counterpart on the other layer that is directly above or below them,
and are referred to as `non-dimer' sites.
Note that some authors \cite{guinea06,nil08,li09,zhang08} employ different
definitions of $A$ and $B$ sites as used here.
The point group of the bilayer crystal structure is
$D_{3d}$ \cite{latil06,manes07,kosh-mcc10}
consisting of elements ($\{ E, 2C_3, 3C_2^{\prime}, i, 2S_6, 3\sigma_d\}$),
and it may be regarded as a direct product of group
$D_3$ ($\{ E, 2C_3, 3C_2^{\prime}\}$)
with the inversion group $C_i$ ($\{ E, i \}$).
Thus, the lattice is symmetric with respect to spatial inversion
symmetry $(x , y , z) \rightarrow (-x , -y , -z)$.

Primitive reciprocal lattice vectors $\mathbf{b}_1$ and $\mathbf{b}_2$ of
monolayer and bilayer graphene, where
$\mathbf{a}_1 \cdot \mathbf{b}_1 = \mathbf{a}_2 \cdot \mathbf{b}_2 = 2 \pi$
and $\mathbf{a}_1 \cdot \mathbf{b}_2 = \mathbf{a}_2 \cdot \mathbf{b}_1 = 0$,
are given by
\begin{eqnarray}
\mathbf{b}_1 = \left( \frac{2\pi}{a} , \frac{2\pi}{\sqrt{3}a} \right) \, , \qquad
\mathbf{b}_2 = \left( \frac{2\pi}{a} , - \frac{2\pi}{\sqrt{3}a} \right) \, . \label{b1b2}
\end{eqnarray}
As shown in figure~\ref{fig:monolattice}(b), the reciprocal lattice is an hexagonal Bravais lattice,
and the first Brillouin zone is an hexagon.

\subsection{The tight-binding model}

\subsubsection{An arbitrary crystal structure}

In the following, we will describe the tight-binding model \cite{ash+mer,saito,mcc12}
and its application to bilayer graphene.
%
We begin by considering an arbitrary crystal with translational
invariance and $M$ atomic orbitals
$\phi_m$ per unit cell, labelled by index $m = 1 \ldots M$.
Bloch states
$\Phi_m ( \mathbf{k} , \mathbf{r} )$
for a given position vector $\mathbf{r}$ and
wave vector $\mathbf{k}$
may be written as
\begin{eqnarray}
\Phi_m ( \mathbf{k} , \mathbf{r} ) = \frac{1}{\sqrt{N}}
\sum_{i=1}^{N} e^{i \mathbf{k}.\mathbf{R}_{m,i}}
\phi_m \left( \mathbf{r} - \mathbf{R}_{m,i} \right) \, , \label{bloch}
\end{eqnarray}
where $N$ is the number of unit cells, $i = 1 \ldots N$ labels the unit cell,
and $\mathbf{R}_{m,i}$ is the position vector of the $m$th orbital in the $i$th unit cell.

The electronic wave function $\Psi_j ( \mathbf{k} , \mathbf{r} )$ may be expressed
as a linear superposition of Bloch states
\begin{eqnarray}
\Psi_j ( \mathbf{k} , \mathbf{r} ) = \sum_{m=1}^{M}
{\psi}_{j,m} ( \mathbf{k} ) \, \Phi_m ( \mathbf{k} , \mathbf{r} ) \, , \label{exp}
\end{eqnarray}
where ${\psi}_{j,m}$ are expansion coefficients.
There are $M$ different energy bands, and the energy $E_j (\mathbf{k})$ of the $j$th band
is given by $E_j (\mathbf{k}) = \langle \Psi_j | {\cal H} | \Psi_j \rangle /\langle \Psi_j | \Psi_j \rangle$
where ${\cal H}$ is the Hamiltonian.
Minimising the energy $E_j$ with respect to the expansion coefficients ${\psi}_{j,m}$
\cite{saito,mcc12}
leads to
\begin{eqnarray}
H \psi_j &=& E_j S \psi_j \, ,   \label{HES}
\end{eqnarray}
where $\psi_j$ is a
column vector, $\psi_j^T = \left( {\psi}_{j1} , {\psi}_{j2} , \ldots , {\psi}_{jM} \right)$.
The transfer integral matrix $H$
and overlap integral matrix $S$ are $M \times M$ matrices with
matrix elements defined by
\begin{eqnarray}
H_{mm'} = \langle \Phi_m | {\cal H} | \Phi_{m'} \rangle \, , \qquad
S_{mm'} = \langle \Phi_m | \Phi_{m'} \rangle \, . \label{mcc}
\end{eqnarray}
The band energies $E_j$ may be determined from the generalised eigenvalue
equation~(\ref{HES}) by solving the secular equation
\begin{eqnarray}
\det \left( H -  E_j S \right) = 0 \, , \label{sec}
\end{eqnarray}
where `$\det$' stands for the determinant of the matrix.

In order to model a given system in terms of the generalised
eigenvalue problem~(\ref{HES}), it is necessary to determine
the matrices $H$ and $S$. We will proceed by considering the relatively
simple case of monolayer graphene, before generalising the approach to bilayers.
In the following sections, we will omit the subscript $j$ on
$\psi_j$ and $E_j$ in equation~(\ref{HES}), remembering
that the number of solutions is $M$, the number of
orbitals per unit cell.

\subsubsection{Monolayer graphene}

Here, we will outline how to apply the tight-binding model to graphene, and refer the
reader to tutorial-style reviews \cite{saito,mcc12} for further details.
We take into account one $2p_z$ orbital per atomic
site and, as there are two atoms in the unit cell of monolayer graphene,
figure~\ref{fig:monolattice}(a),
we include two orbitals per unit cell labelled as
 $m = A$ and $m = B$
(the $A$ atoms and the $B$ atoms
are each arranged on an hexagonal Bravais lattice).

We begin by considering the diagonal element $H_{AA}$ of the transfer integral matrix
$H$, equation~(\ref{mcc}), for the $A$ site orbital.
It may be determined by substituting the Bloch function~(\ref{bloch}) for $m=A$ into the
matrix element~(\ref{mcc}), which results in a double sum over the positions of the unit
cells in the crystal. Assuming that the dominant contribution arises from those
terms involving a given orbital interacting with itself ({\em i.e.}, in the same unit cell),
the matrix element may be written as
\begin{eqnarray}
H_{AA} \approx \frac{1}{N} \sum_{i=1}^{N}
\langle \phi_A \left( \mathbf{r} - \mathbf{R}_{A,i} \right) | {\cal H} | \phi_A \left( \mathbf{r} - \mathbf{R}_{A,i} \right) \rangle \, .
\end{eqnarray}
This may be regarded as a summation over all unit cells of a parameter
$\epsilon_{A} =
\langle \phi_A \left( \mathbf{r} - \mathbf{R}_{A,i} \right) | {\cal H} | \phi_A \left( \mathbf{r} - \mathbf{R}_{A,i} \right) \rangle$ that takes the same value in every unit cell.
Thus, the matrix element may
be simply expressed as $H_{AA} \approx \epsilon_{A}$.
Similarly, the diagonal element $H_{BB}$
for the $B$ site orbital can be written as $H_{BB} = \epsilon_B$,
while for intrinsic graphene  $\epsilon_A$ is equal to $\epsilon_B$
as the two sublattices are identical.
The calculation of the diagonal elements of the overlap integral matrix
$S$, equation~(\ref{mcc}), proceeds in the same way as that of $H$, with
the overlap of an orbital with itself equal to unity,
$\langle \phi_j \left( \mathbf{r} - \mathbf{R}_{j,i} \right) | \phi_j \left( \mathbf{r} - \mathbf{R}_{j,i} \right) \rangle = 1$. Thus, $S_{BB} = S_{AA} = 1$.

The off-diagonal element $H_{AB}$ of the transfer integral matrix
$H$ describes the possibility of hopping between orbitals on $A$ and $B$ sites.
Substituting the Bloch function~(\ref{bloch}) into the matrix element~(\ref{mcc}) results
in a sum over all $A$ sites and a sum over all $B$ sites. We assume that the dominant
contribution arises from hopping between adjacent sites.
If we consider a given $A$ site, say,
then we take into account the possibility of hopping
to its three nearest-neighbour $B$ sites,
labelled by index $l = 1,2,3$:
\begin{eqnarray}
H_{AB} &\approx& \frac{1}{N} \sum_{i=1}^{N} \sum_{l=1}^{3}
e^{i \mathbf{k}. \mbox{\boldmath$\delta$}_l} \nonumber \\
&& \times \langle \phi_A
\left( \mathbf{r} - \mathbf{R}_{A,i}
\right) | {\cal H} |
\phi_B \left( \mathbf{r} - \mathbf{R}_{A,i}
-   \mbox{\boldmath$\delta$}_l
\right) \rangle \, ,
\end{eqnarray}
where $\mbox{\boldmath$\delta$}_l$ are the positions of three nearest $B$ atoms
relative to a given $A$ atom, which may be written as
$\mbox{\boldmath$\delta$}_1 = \left( 0 , a/\sqrt{3}\right)$,
$\mbox{\boldmath$\delta$}_2 = \left( a/2 , -a/{2\sqrt{3}}\right)$,
$\mbox{\boldmath$\delta$}_3 = \left( - a/2 , -{a}/{2\sqrt{3}}\right)$.

The sum with respect to the three nearest-neighbour $B$ sites is identical for every $A$ site.
A hopping parameter may be defined as
\begin{equation}
\gamma_0 = -
\langle \phi_A \left( \mathbf{r} - \mathbf{R}_{A,i} \right) | {\cal H} |
\phi_B \left( \mathbf{r} - \mathbf{R}_{A,i} - \mbox{\boldmath$\delta$}_l \right) \rangle,
\end{equation}
which is positive.
Then, the matrix element may be written as
\begin{eqnarray}
H_{AB} \approx - \gamma_0 f \left( \mathbf{k} \right) \, ; \quad
f \left( \mathbf{k} \right) &=& \sum_{l=1}^{3}
e^{i \mathbf{k}. \mbox{\boldmath$\delta$}_l} \, ,
\label{hab}
\end{eqnarray}
The other off-diagonal matrix element is given by
$H_{BA} = H_{AB}^{\ast} \approx - \gamma_0 f^{\ast} \left( \mathbf{k} \right)$.
The function $f \left( \mathbf{k} \right)$ describing
nearest-neighbor hopping, equation~(\ref{hab}), is given by
\begin{eqnarray}
f \left( \mathbf{k} \right) = e^{i k_y a /\sqrt{3}} + 2 e^{- i k_y a /2\sqrt{3}} \cos \left( k_x a / 2 \right) \, , \label{fk}
\end{eqnarray}
where $\mathbf{k} = \left( k_x , k_y \right)$ is the in-plane wave vector.
The calculation of the off-diagonal elements of the overlap integral matrix $S$ is similar to
those of $H$.
A parameter $s_0 = \langle \phi_A \left( \mathbf{r} - \mathbf{R}_{A,i} \right) | \phi_B \left( \mathbf{r} - \mathbf{R}_{B,l} \right) \rangle$ is introduced to describe the possibility of non-zero
overlap between orbitals on adjacent sites, giving
$S_{AB} = S_{BA}^{\ast} = s_0 f \left( \mathbf{k} \right)$.

Gathering the matrix elements, the transfer $H_{\mathrm{m}}$ and
overlap $S_{\mathrm{m}}$ integral matrices of monolayer
graphene may be written as
\begin{eqnarray}
H_{\mathrm{m}} &=& \left(
      \begin{array}{cc}
        \epsilon_{A} & - \gamma_0 f \left( \mathbf{k} \right) \\
        - \gamma_0 f^{\ast} \left( \mathbf{k} \right) & \epsilon_{B} \\
      \end{array}
    \right) \, , \label{Hmono} \\
S_{\mathrm{m}} &=& \left(
      \begin{array}{cc}
        1 & s_0 f \left( \mathbf{k} \right) \\
        s_0 f^{\ast} \left( \mathbf{k} \right) & 1 \\
      \end{array}
    \right) \, . \label{Smono}
\end{eqnarray}
The corresponding energy may be determined \cite{saito} by solving the
secular equation~(\ref{sec}). For intrinsic graphene,
{\em i.e.}, $\epsilon_A = \epsilon_B = 0$, we have
\begin{eqnarray}
E_{\pm} =  \frac{\pm \gamma_0 |f \left( \mathbf{k} \right) |}{1 \mp s_0 |f \left( \mathbf{k} \right) |} \, . \label{monofull}
\end{eqnarray}
The parameter values are listed by Saito {\em et al} \cite{saito}
as $\gamma_0 = 3.033\,$eV and $s_0 = 0.129$.

The function $f ( \mathbf{k} )$, equation~(\ref{fk}) is zero at the corners of
the Brillouin zone, two of which are non-equivalent ({\em i.e.}, they are
not connected by a reciprocal lattice vector). For example, corners
$K_{+}$ and $K_{-}$ with wave vectors $\mathbf{K}_{\pm} = \pm(4\pi/3a , 0)$
are labelled in Figure~\ref{fig:monolattice}(b).
Such positions are called $K$ points or valleys, and we will use a valley
index $\xi = \pm 1$ to distinguish points $K_{\xi}$.
At these positions, the solutions~(\ref{monofull}) are degenerate,
marking a crossing point and
zero band gap between the conduction and valence bands.
The transfer matrix $H_{\mathrm{m}}$ is approximately
equal to a Dirac-like Hamiltonian in the vicinity of the $K$ point,
describing massless chiral quasiparticles with a linear dispersion relation.
These points are particularly important because the Fermi level is located
near them in pristine graphene.

\subsubsection{Bilayer graphene}

In the tight-binding description of bilayer graphene, we take into
account $2p_z$ orbitals on the four atomic sites in the unit
cell, labelled as $j = A1,B1,A2,B2$. Then, the transfer
integral matrix of bilayer graphene
\cite{mcc06a,guinea06,part06,mucha08,nil08,mucha10}
is a $4 \times 4$ matrix given by
\begin{eqnarray}
H_{\mathrm{b}} = \left(
      \begin{array}{cccc}
        \epsilon_{A1} & - \gamma_0 f \left( \mathbf{k} \right) & \gamma_4 f \left( \mathbf{k} \right) & -\gamma_3 f^{\ast} \left( \mathbf{k} \right) \\
        - \gamma_0 f^{\ast} \left( \mathbf{k} \right) & \epsilon_{B1} & \gamma_1 & \gamma_4 f \left( \mathbf{k} \right) \\
        \gamma_4 f^{\ast} \left( \mathbf{k} \right) & \gamma_1 & \epsilon_{A2} & - \gamma_0 f \left( \mathbf{k} \right) \\
        -\gamma_3 f \left( \mathbf{k} \right) & \gamma_4 f^{\ast} \left( \mathbf{k} \right) & - \gamma_0 f^{\ast} \left( \mathbf{k} \right) & \epsilon_{B2} \\
      \end{array}
    \right)  , \nonumber\\ \label{Hbfull}
\end{eqnarray}
where the tight-binding parameters are defined as
\begin{eqnarray}
\gamma_0 &=& - \langle \phi_{A1} | {\cal H} | \phi_{B1} \rangle = - \langle \phi_{A2} | {\cal H} | \phi_{B2} \rangle , \\
\gamma_1 &=& \langle \phi_{A2} | {\cal H} | \phi_{B1} \rangle , \\
\gamma_3 &=& - \langle \phi_{A1} | {\cal H} | \phi_{B2} \rangle , \\
\gamma_4 &=& \langle \phi_{A1} | {\cal H} | \phi_{A2} \rangle = \langle \phi_{B1} | {\cal H} | \phi_{B2} \rangle .
\end{eqnarray}
Here, we use the notation of the Slonczewski-Weiss-McClure (SWM) model
\cite{sw58,mcclure57,mcclure60,dressel02} that describes bulk graphite.
Note that definitions of the parameters used by authors can differ,
particularly with respect to signs.

\begin{table*}[tbp]
\caption{Values (in eV) of the Slonczewski-Weiss-McClure (SWM) model
parameters \cite{sw58,mcclure57,mcclure60,dressel02} determined experimentally.
Numbers in parenthesis indicate estimated accuracy of the final digit(s).
The energy difference between dimer and non-dimer sites
in the bilayer is $\Delta^{\prime} = \Delta - \gamma_2 + \gamma_5$.
Note that next-nearest layer parameters $\gamma_2$ and $\gamma_5$ are
not present in bilayer graphene.}\label{tab:1}
\renewcommand{\tabcolsep}{0.4cm}
\begin{tabular}{lllllll}
  \hline
  \hline
  Parameter & Graphite \cite{dressel02} & Bilayer \cite{mal07} & Bilayer \cite{zhang08} & Bilayer \cite{li09} & Bilayer \cite{kuz09b} & Trilayer \cite{tay11}\\
\hline
  $\gamma_0$ & 3.16(5) & 2.9 & 3.0\footnotemark[1] & - & 3.16(3) & 3.1\footnotemark[1] \\
  $\gamma_1$ & 0.39(1) & 0.30 & 0.40(1) & 0.404(10) & 0.381(3)  & 0.39\footnotemark[1] \\
  $\gamma_2$ & -0.020(2) & - & - & - & - & -0.028(4) \\
  $\gamma_3$ & 0.315(15) & 0.10 & 0.3\footnotemark[1] & - & 0.38(6) & 0.315\footnotemark[1] \\
  $\gamma_4$ & 0.044(24) & 0.12 & 0.15(4) & - & 0.14(3) & 0.041(10) \\
  $\gamma_5$ & 0.038(5) & - & - & - & - & 0.05(2) \\
  $\Delta$ & -0.008(2) & - & 0.018(3) & 0.018(2) & 0.022(3) & -0.03(2) \\
  $\Delta^{\prime}$ & 0.050(6) & - & 0.018(3) & 0.018(2) & 0.022(3) & 0.046(10) \\
  \hline
  \hline
\footnotetext[1]{This parameter was not determined by the given experiment, the value quoted was taken
from previous literature.}
\end{tabular}
\end{table*}

The upper-left and lower-right square $2 \times 2$ blocks of
$H_{\mathrm{b}}$ describe intra-layer terms
and are simple generalisations of the monolayer,
equation~(\ref{Hmono}). For bilayer graphene,
however, we include parameters
describing the on-site energies $\epsilon_{A1}$, $\epsilon_{B1}$,
$\epsilon_{A2}$, $\epsilon_{B2}$ on the four atomic sites,
that are not equal in the most general case.
As there are four sites, differences between them are described by three parameters \cite{mucha10}:
\begin{eqnarray}
\epsilon_{A1} &=& {\textstyle\frac{1}{2}}\left(-U + \delta_{AB} \right) \, , \\
\epsilon_{B1} &=& {\textstyle\frac{1}{2}}\left(-U + 2\Delta^{\prime} - \delta_{AB} \right) \, , \\
\epsilon_{A2} &=& {\textstyle\frac{1}{2}}\left(U + 2\Delta^{\prime} + \delta_{AB} \right) \, , \\
\epsilon_{B2} &=& {\textstyle\frac{1}{2}}\left(U - \delta_{AB} \right) \, ,
\end{eqnarray}
where
\begin{eqnarray}
U &=& {\textstyle\frac{1}{2}}\left[(\epsilon_{A1} + \epsilon_{B1}) - (\epsilon_{A2} +
\epsilon_{B2})\right] \, , \label{defu} \\
\Delta^{\prime} &=& {\textstyle\frac{1}{2}}\left[(\epsilon_{B1} +
\epsilon_{A2}) - (\epsilon_{A1} + \epsilon_{B2})\right] \, , \label{defd} \\
\delta_{AB} &=& {\textstyle\frac{1}{2}}\left[(\epsilon_{A1} + \epsilon_{A2}) - (\epsilon_{B1} +
\epsilon_{B2})\right] \, . \label{defdab}
\end{eqnarray}
The three independent parameters are $U$ to describe interlayer asymmetry between the two layers
\cite{mcc06a,ohta06,guinea06,mcc06b,min07,castro,oostinga,aoki07,mcc07,guinea07,mcc07b,guinea07b,gava,bouk08},
$\Delta^{\prime}$ for an energy difference between dimer and non-dimer sites \cite{dressel02,nil08,zhang08,li09},
and $\delta_{AB}$ for an energy difference between $A$ and $B$ sites on each
layer \cite{mucha09d, mucha10}.
These parameters are described in detail in sections~\ref{s:gfour} and \ref{s:endif}.

The upper-right and lower-left square $2 \times 2$ blocks of
$H_{\mathrm{b}}$ describe inter-layer coupling.
Parameter $\gamma_1$ describes coupling between pairs of orbitals on
the dimer sites $B1$ and $A2$: since this is a vertical coupling,
the corresponding terms in $H_{\mathrm{b}}$
({\em i.e.}, $H_{A2,B1} = H_{B1,A2} = \gamma_1$)
do not contain $f \left( \mathbf{k} \right)$ which
describes in-plane hopping.
Parameter $\gamma_3$ describes interlayer coupling between non-dimer
orbitals $A1$ and $B2$, and $\gamma_4$ describes interlayer coupling between
dimer and non-dimer orbitals $A1$ and $A2$ or $B1$ and $B2$.
Both $\gamma_3$ and $\gamma_4$ couplings are `skew': they are not strictly vertical,
but involve a component of in-plane hopping, and each atom on one
layer ({\em e.g.}, $A1$ for $\gamma_3$) has
three equidistant nearest-neighbours ({\em e.g.}, $B2$ for $\gamma_3$) on the other layer.
In fact, the in-plane component of this skew hopping is analogous to nearest-neighbour
hopping within a single layer, as parameterised by $\gamma_0$.
Hence, the skew interlayer hopping ({\em e.g.}, $H_{A1,B2} = -\gamma_3 f^{\ast} \left( \mathbf{k} \right)$)
contains the factor $f \left( \mathbf{k} \right)$ describing in-plane hopping.

It is possible to introduce an overlap integral matrix for bilayer graphene \cite{mucha10}
\begin{eqnarray}
S_{\mathrm{b}} = \left(
      \begin{array}{cccc}
        1 & s_0 f \left( \mathbf{k} \right) & 0 & 0 \\
        s_0 f^{\ast} \left( \mathbf{k} \right) & 1 & s_1 & 0 \\
        0 & s_1 & 1 & s_0 f \left( \mathbf{k} \right) \\
        0 & 0 & s_0 f^{\ast} \left( \mathbf{k} \right) & 1 \\
      \end{array}
    \right) \,  , \label{Sbfull}
\end{eqnarray}
with a form that mirrors $H_{\mathrm{b}}$.
Here, we only include two parameters:
$s_0 = \langle \phi_{A1} | \phi_{B1} \rangle = \langle \phi_{A2} | \phi_{B2} \rangle$
describing non-orthogonality of intra-layer nearest-neighbours and
$s_1 = \langle \phi_{A2} | \phi_{B1} \rangle$ describing non-orthogonality of orbitals
on dimer sites $A1$ and $B2$. In principle, it is possible to introduce additional
parameters analogous to $\gamma_3$, $\gamma_4$, {\em etc.}, but generally they will be small
and irrelevant.
In fact, it is common practice to neglect the overlap integral matrix entirely,
{\em i.e.}, replace $S_{\mathrm{b}}$ with a unit matrix, because the influence of
parameters $s_0$ and $s_1$ describing non-orthogonality of adjacent orbitals is small
at low energy $|E| \leq \gamma_1$. Then, the generalised eigenvalue
equation~(\ref{HES}) reduces to an eigenvalue equation
$H_{\mathrm{b}} \psi = E \psi$ with Hamiltonian $H_{\mathrm{b}}$,
equation~(\ref{Hbfull}).

The energy differences $U$ and $\delta_{AB}$ are usually attributed to extrinsic
factors such as gates, substrates or doping. Thus, there are five independent
parameters in the Hamiltonian~(\ref{Hbfull}) of intrinsic bilayer graphene, namely
$\gamma_0$, $\gamma_1$, $\gamma_3$, $\gamma_4$ and $\Delta^{\prime}$.
The band structure predicted by the tight-binding model has been compared
to observations from photoemission \cite{ohta06}, Raman \cite{mal07}
and infrared spectroscopy \cite{hen08,zhang08,li09,kuz09a,kuz09b,mak10}.
Parameter values determined by fitting to experiments are listed in Table~\ref{tab:1}
for bulk graphite \cite{dressel02}, for bilayer graphene by Raman \cite{mal07,das}
and infrared \cite{zhang08,li09,kuz09b} spectroscopy, and for Bernal-stacked trilayer graphene
by observation of Landau level crossings \cite{tay11}. Note that
there are seven parameters in the Slonczewski-Weiss-McClure (SWM) model of graphite
\cite{sw58,mcclure57,mcclure60,dressel02}
because the next-nearest layer couplings $\gamma_2$ and $\gamma_5$, absent in bilayer,
are present in graphite (and trilayer graphene, too).
Parameter $\Delta$ in the SWM model is related by
$\Delta = \Delta^{\prime} + \gamma_2 - \gamma_5$ to the parameter $\Delta^{\prime}$
describing the energy difference between dimer and non-dimer sites in bilayer
graphene.

\begin{figure}[t]
\centerline{\epsfxsize=0.7\hsize \epsffile{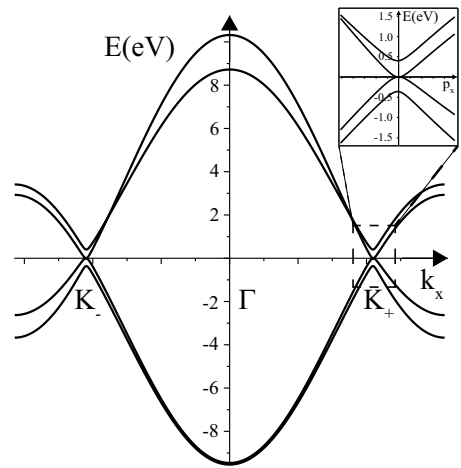}}
\caption{Low-energy bands of bilayer graphene arising from $2p_z$ orbitals plotted along the $k_x$
axis in reciprocal space intersecting the corners $K_{-}$, $K_{+}$
and the centre $\Gamma$ of the Brillouin zone. The inset shows the bands in
the vicinity of the $K_{+}$ point. Plots were made using
Hamiltonian $H_{\mathrm{b}}$, equation~(\ref{Hbfull}),
with parameter values $\gamma_0=3.16\,$eV, $\gamma_1=0.381\,$eV,
$\gamma_3=0.38\,$eV, $\gamma_4=0.14\,$eV,
$\epsilon_{B1} = \epsilon_{A2} = \Delta^{\prime}=0.022\,$eV,
and $\epsilon_{A1} = \epsilon_{B2} = U = \delta_{AB}=0$ \cite{kuz09b}.}
\label{fig:bands}
\end{figure}

The energy bands are plotted in figure~\ref{fig:bands} along the $k_x$
axis in reciprocal space intersecting the corners $K_{-}$, $K_{+}$
and the centre $\Gamma$ of the Brillouin zone [see figure~\ref{fig:monolattice}(b)].
Plots were made using Hamiltonian $H_{\mathrm{b}}$, equation~(\ref{Hbfull}),
with parameter values determined by infrared spectroscopy $\gamma_0=3.16\,$eV, $\gamma_1=0.381\,$eV,
$\gamma_3=0.38\,$eV, $\gamma_4=0.14\,$eV,
$\epsilon_{B1} = \epsilon_{A2} = \Delta^{\prime}=0.022\,$eV,
and $\epsilon_{A1} = \epsilon_{B2} = U = \delta_{AB}=0$ \cite{kuz09b}.
There are four bands because the model takes into account one $2p_z$ orbital
on each of the four atomic sites in the unit cell; a pair of conduction bands
and a pair of valence bands. Over most of the Brillouin zone, each pair is split by an energy
of the order of the interlayer spacing $\gamma_1 \approx 0.4\,$eV \cite{trickey}.
Near the $K$ points, inset of figure~\ref{fig:bands},
one conduction band and one valence band are split away from zero energy by an energy of
the order of the interlayer coupling $\gamma_1$, whereas two bands touch
at zero energy \cite{mcc06a}. The `split' bands are a bonding and anti-bonding
pair arising from the strong coupling (by interlayer coupling $\gamma_1$) of the
orbitals on the dimer $B1$ and $A2$ sites, whereas the `low-energy' bands
arise from hopping between the non-dimer $A1$ and $B2$ sites.
In pristine bilayer graphene, the Fermi level lies at the point where the
two low-energy bands touch (shown as zero energy in figure~\ref{fig:bands})
and, thus, this region is relevant for the study of electronic properties.
It will be the focus of the following sections.

At low energy, the shape of the band structure predicted by the tight-binding model
(see inset in figure~\ref{fig:bands}) is in good agreement with that calculated by
density functional theory \cite{latil06,min07,aoki07} and it is possible
obtain values for the tight-binding parameters in this way, generally
in line with the experimental ones listed in Table~\ref{tab:1}.
The tight-binding model Hamiltonian $H_{\mathrm{b}}$, equation~(\ref{Hbfull}),
used in conjuction with the parameters listed in Table~\ref{tab:1}, is not accurate
over the whole Brillouin zone because the fitting of tight-binding parameters
is generally done in the vicinity of the corners of the Brillouin zone $K_{+}$ and
$K_{-}$ (as the Fermi level lies near zero energy).
For example, parameter $s_0$ in equation~(\ref{Sbfull}) describing non-orthogonality of adjacent
orbitals has been neglected here, but it contributes electron-hole
asymmetry which is particularly prevalent near the $\Gamma$ point at the
centre of the Brillouin zone \cite{saito,mcc12}.

\subsubsection{Effective four-band Hamiltonian near the $K$ points}

To describe the properties of electrons in the vicinity of the $K$ points,
a momentum $\mathbf{p} = \hbar \mathbf{k} - \hbar \mathbf{K}_{\xi}$ is introduced which
is measured from the centre of the $K_{\xi}$ point.
Expanding in powers of $\mathbf{p}$, the function $f ( \mathbf{k} )$, equation~(\ref{fk}), is
approximately given by $f ( \mathbf{k} )
\approx - \sqrt{3} a ( \xi p_x - i p_y ) /2 \hbar$
which is valid close to the $K_{\xi}$ point, {\em i.e.}, for $p a / \hbar \ll 1$,
where $p = |\mathbf{p}| = ( p_x^2 + p_y^2 )^{1/2}$.
In monolayer graphene,
the Hamiltonian matrix~(\ref{Hmono})
is simplified by keeping only linear terms in momentum $p$ as
\begin{eqnarray}
H_{\mathrm{m}} &=& \left(
      \begin{array}{cccc}
        \epsilon_{A} & v \pi^{\dagger} \\
        v \pi & \epsilon_B
      \end{array}
    \right) ,
\end{eqnarray}
where $\pi = \xi p_x + i p_y$, $\pi^{\dagger} = \xi p_x - i p_y$,
and $v = \sqrt{3} a \gamma_0 / 2\hbar$ is the band velocity.
In the intrinsic case, $\epsilon_A = \epsilon_B =0$,
the eigen energy becomes $E = \pm v |\mathbf{p}|$,
which approximates Eq.\ (\ref{monofull}).
In bilayer graphene, similarly, Eq.\ (\ref{Hbfull}) is reduced to
\begin{eqnarray}
H_{\mathrm{b}} &=& \left(
      \begin{array}{cccc}
        \epsilon_{A1} & v \pi^{\dagger} & -v_4 \pi^{\dagger} & v_3 \pi \\
        v \pi & \epsilon_{B1} & \gamma_1 & -v_4 \pi^{\dagger}  \\
        -v_4 \pi & \gamma_1 & \epsilon_{A2} & v \pi^{\dagger} \\
        v_3 \pi^{\dagger} & -v_4 \pi & v \pi & \epsilon_{B2} \\
      \end{array}
    \right)  \label{Hbmin} ,
\end{eqnarray}
where we introduced the effective velocities,
$v_3 = \sqrt{3} a \gamma_3 / 2\hbar$ and $v_4 = \sqrt{3} a \gamma_4 / 2\hbar$.

At zero magnetic field Hamiltonian~(\ref{Hbmin}) yields four valley-degenerate bands
$E (\mathbf{p})$. A simple analytic solution may be obtained by neglecting
the terms $v_4 \pi$, $v_4 \pi^{\dagger}$ proportional to $\gamma_4$,
and by considering only interlayer asymmetry $U$ in the on-site
energies: $\epsilon_{A1} = \epsilon_{B1} = -U/2$ and
$\epsilon_{A2} = \epsilon_{B2} = U/2$.
Then, there is electron-hole symmetry, {\em i.e.}, energies may be written
$E = \pm \varepsilon_{\alpha}(\mathbf{p})$, $\alpha =1,2$, \cite{mcc06a} with
\begin{eqnarray}
\varepsilon_{\alpha }^{2} &=& \frac{\gamma_{1}^{2}}{2}+\frac{{U}^{2}}{4}
+\left( v^{2}+\frac{v_{3}^{2}}{2}\right) p^{2}+\left( -1\right)^{\alpha}
\sqrt{\Gamma} \, , \label{efour} \\
\Gamma &=& {\textstyle\frac{1}{4}}\left( \gamma_{1}^{2} - v_{3}^{2}p^{2}\right)^{2}
+ v^{2}p^{2} \left[ \gamma_{1}^{2}+{U}^{2}+v_{3}^{2}p^{2}\right] \nonumber \\
&& \qquad 
+ \,2\xi \gamma _{1}v_{3}v^{2}p^{3}\cos 3\varphi \, , \nonumber
\end{eqnarray}
where $\varphi$ is the polar angle of momentum
$\mathbf{p} = \left( p_x , p_y \right) = p \left( \cos \varphi , \sin \varphi \right)$.
Energy $\varepsilon_{2}$ describes the higher-energy bands split from
zero energy by the interlayer coupling $\gamma_1$
between the orbitals on the dimer sites $B1$, $A2$.

Low-energy bands $E = \pm \varepsilon_{1}$ are related to orbitals on the non-dimer sites
$A1$, $B2$. In an intermediate energy range
$U , (v_3/v)^2 \gamma_1 < \varepsilon_{1} < \gamma_1$
it is possible to neglect the interlayer asymmetry $U$
and terms proportional to $\gamma_3$ ({\em i.e.}, set $U = v_3 = 0$),
and the low-energy bands may be approximated \cite{mcc06a} as
\begin{equation}
\varepsilon_{1} \approx {\textstyle\frac{1}{2}}\gamma _{1}\left[
\sqrt{1+4v^{2}p^{2}/\gamma _{1}^{2}}-1\right] ,  \label{epone}
\end{equation}
which interpolates between an approximately linear dispersion
$\varepsilon_{1} \approx v p$ at large momentum
to a quadratic one $\varepsilon_{1} \approx p^2/2m$ at small
momentum, where the mass is $m = \gamma_1/2v^2$ (see inset in figure~\ref{fig:bands}).
This crossover occurs at $p \approx \gamma_1 /2v$.
A convenient way to describe the bilayer at low energy and momentum
$p \ll \gamma_1 /2v$ is to eliminate the components in the Hamiltonian~(\ref{Hbmin})
related to the orbitals on dimer sites $B1$, $A2$,
resulting in an effective two-component Hamiltonian describing
the orbitals on the non-dimer sites $A1$, $B2$, and, thus, the two
bands that approach each other at zero energy.
This is described in the next section, and the solutions of this Hamiltonian
are shown to be massive chiral quasiparticles \cite{novo06,mcc06a}, as opposed to massless chiral ones in
monolayer graphene.

\subsection{Effective two-band Hamiltonian at low energy}

In this section we focus on the low-energy electronic band structure
in the vicinity of the points $K_{+}$ and $K_{-}$ at the corners of
the first Brillouin zone, relevant for energies near the Fermi level.
A simple model may be obtained by eliminating orbitals related
to the dimer sites, resulting in an effective Hamiltonian for the
low-energy orbitals. First, we outline the procedure in general terms,
because it may be applied to systems other than bilayer graphene
such as $ABC$-stacked (rhombohedral) graphene multilayers \cite{koshinoABC,zhang10},
before applying it specifically to bilayer graphene.

\subsubsection{General procedure}

We consider the energy eigenvalue equation, and consider separate
blocks in the Hamiltonian corresponding to
low-energy $\theta = \left( \psi_{A1} , \psi_{B2} \right)^T$
and dimer $\chi = \left( \psi_{A2} , \psi_{B1} \right)^T$ components:
\begin{eqnarray}
\left(
  \begin{array}{cc}
    h_{\theta} & u \\
    u^{\dagger} & h_{\chi} \\
  \end{array}
\right)
\left(
  \begin{array}{c}
    \theta \\
    \chi \\
  \end{array}
\right)
= E \left(
  \begin{array}{c}
    \theta \\
    \chi \\
  \end{array}
\right) \, , \label{tc}
\end{eqnarray}
The second row of~(\ref{tc}) allows the dimer components to be expressed in terms of
the low-energy ones:
\begin{eqnarray}
\chi = \left( E - h_{\chi} \right)^{-1} u^{\dagger} \theta \, , \label{high}
\end{eqnarray}
Substituting this into the first row of~(\ref{tc}) gives an effective eigenvalue
equation written solely for the low-energy components:
\begin{eqnarray*}
\left[ h_{\theta} + u \left( E - h_{\chi} \right)^{-1} u^{\dagger} \right] \theta &=& E \theta \, , \\
\left[ h_{\theta} - u h_{\chi}^{-1} u^{\dagger} \right] \theta &\approx& E {\cal S} \theta \, ,
\end{eqnarray*}
where ${\cal S} = 1 + u h_{\chi}^{-2} u^{\dagger}$. The second equation is accurate up to linear terms in $E$.
Finally, we perform a transformation $\Phi = {\cal S}^{1/2} \theta$:
\begin{eqnarray}
\left[ h_{\theta} - u h_{\chi}^{-1} u^{\dagger} \right] {\cal S}^{-1/2} \Phi &\approx& E {\cal S}^{1/2} \Phi \, , \nonumber \\
{\cal S}^{-1/2} \left[ h_{\theta} - u h_{\chi}^{-1} u^{\dagger} \right] {\cal S}^{-1/2} \Phi &\approx& E \Phi \, . \label{heff}
\end{eqnarray}
This transformation ensures that normalisation of $\Phi$ is consistent with that of the original states:
\begin{eqnarray*}
\Phi^{\dagger}  \Phi = \theta^{\dagger} {\cal S} \theta
&=& \theta^{\dagger} \left( 1 + u h_{\chi}^{-2} u^{\dagger} \right) \theta \, , \\
&\approx& \theta^{\dagger} \theta + \chi^{\dagger} \chi \, ,
\end{eqnarray*}
where we used equation~(\ref{high}) for small $E$:
$\chi \approx - h_{\chi}^{-1} u^{\dagger} \theta$.
Thus, the effective Hamiltonian for low-energy components is given by equation~(\ref{heff}):
\begin{eqnarray}
H^{(\mathrm{eff})} &\approx& {\cal S}^{-1/2}
\left[ h_{\theta} - u h_{\chi}^{-1} u^{\dagger} \right] {\cal S}^{-1/2} \, , \label{trans1} \\
{\cal S} &=& 1 + u h_{\chi}^{-2} u^{\dagger} \, . \label{trans2}
\end{eqnarray}

\subsubsection{Bilayer graphene}

The Hamiltonian~(\ref{Hbmin}) is written in basis $A1,B1,A2,B2$.
If, instead, it is written in the basis of low-energy and dimer
components $( \theta, \chi ) \equiv A1,B2,A2,B1$, equation~(\ref{tc}), then
\begin{eqnarray*}
h_{\theta} &=& \left(
               \begin{array}{cc}
                 \epsilon_{A1} & v_3\pi \\
                 v_3\pi^{\dagger} & \epsilon_{B2} \\
               \end{array}
             \right) \! , \qquad
h_{\chi} = \left(
               \begin{array}{cc}
                 \epsilon_{A2} & \gamma_1 \\
                 \gamma_1 & \epsilon_{B1} \\
               \end{array}
             \right) \! , \\
u &=& \left(
               \begin{array}{cc}
                 -v_4\pi^{\dagger} & v\pi^{\dagger} \\
                 v\pi & -v_4\pi \\
               \end{array}
             \right) \! , \qquad
u^{\dagger} = \left(
               \begin{array}{cc}
                 -v_4\pi & v\pi^{\dagger} \\
                 v\pi & -v_4\pi^{\dagger} \\
               \end{array}
             \right) \! .
\end{eqnarray*}
Using the procedure described in the previous section,
equations~(\ref{trans1},\ref{trans2}), it is possible to
obtain an effective Hamiltonian $H^{(\mathrm{eff})} \equiv {\hat{H}}_{2}$
for components $(\psi_{A1},\psi_{B2})$.
An expansion is performed by assuming that the
intralayer hopping $\gamma_0$ and the interlayer coupling
$\gamma_1$ are larger than other energies:
$\gamma_{0}, \gamma_{1} \gg |E|, vp, |\gamma_3 |,
|\gamma_4 |, |U|, |\Delta^{\prime}|, |\delta_{AB}|$.
Then, keeping only terms that are linear in the small parameters
$|\gamma_3 |, |\gamma_4 |, |U|, |\Delta^{\prime}|, |\delta_{AB}|$ and quadratic in momentum,
the effective Hamiltonian \cite{mcc06a,mucha10} is
\begin{eqnarray}
{\hat{H}}_{2} &=& {\hat{h}}_{0} + {\hat{h}}_{w} + {\hat{h}}_{4}
+ {\hat{h}}_{\Delta} + {\hat{h}}_{U} + {\hat{h}}_{AB} ,  \label{heff1} \\
{\hat{h}}_{0} &=&-\frac{1}{2m}\left(
\begin{array}{cc}
0 & \left( {\pi }^{\dag }\right) ^{2} \\
{\pi ^{2}} & 0
\end{array} \right) , \nonumber \\
{\hat{h}}_{w} &=& v_{3}\left(
\begin{array}{cc}
0 & {\pi } \\
{\pi }^{\dag } & 0
\end{array}
\right) -  \frac{v_3 a}{4\sqrt{3}\hbar} \left( \begin{array}{cc}
0 & \left( {\pi }^{\dag }\right) ^{2} \\
{\pi ^{2}} & 0
\end{array} \right) , \nonumber
\\
{\hat{h}}_{4} &=& \frac{2v v_4}{\gamma_1} \left(
\begin{array}{cc}
\pi^{\dagger}\pi & 0 \\
0 & \pi \pi^{\dagger}
\end{array}
\right)  ,  \nonumber
\\
{\hat{h}}_{\Delta} &=& \frac{\Delta^{\prime}v^2}{\gamma_1^2} \left(
\begin{array}{cc}
\pi^{\dagger}\pi & 0 \\
0 & \pi \pi^{\dagger}
\end{array}
\right)  ,  \nonumber
\\
{\hat{h}}_{U} &=& - \frac{U}{2} \!\! \left[ \left(
\begin{array}{cc}
1 & 0 \\
0 & -1
\end{array}
\right)
- \frac{2v^2}{\gamma_1^2}
\left(
  \begin{array}{cc}
\pi^{\dagger}\pi & 0 \\
    0 & -\pi \pi^{\dagger} \\
  \end{array}
\right) \right]  ,  \nonumber
\\
{\hat{h}}_{AB} &=& \frac{\delta_{AB}}{2} \left(
\begin{array}{cc}
1 & 0 \\
0 & -1
\end{array}
\right) ,  \nonumber
\end{eqnarray}
where $\pi = \xi p_x + i p_y$, $\pi^{\dagger} = \xi p_x - i p_y$.
In the following sections, we discuss the terms in ${\hat{H}}_{2}$.
The first term ${\hat{h}}_{0}$ describes massive chiral electrons, section~\ref{s:mce}.
It generally dominates at low energy $|E| \ll \gamma_1$, so that the other terms
in ${\hat{H}}_{2}$ may be considered as perturbations of it.
The second term ${\hat{h}}_{w}$, section~\ref{s:tw}, introduces a triangular distortion of
the Fermi circle around each $K$ point known as `trigonal warping'.
Terms ${\hat{h}}_{U}$ and ${\hat{h}}_{AB}$, with $\pm 1$ on the diagonal, produce
a band gap between the conduction and valence bands, section~\ref{s:endif},
whereas ${\hat{h}}_{4}$ and ${\hat{h}}_{\Delta}$ introduce electron-hole
asymmetry into the band structure, section~\ref{s:gfour}.

The Hamiltonian~(\ref{heff1}) is written in the vicinity of a valley
with index $\xi = \pm 1$ distinguishing between $K_{+}$ and $K_{-}$.
In order to briefly discuss the effect of symmetry operations on it,
we introduce Pauli spin matrices $\sigma_x$, $\sigma_y$, $\sigma_z$
in the $A1$/$B2$ sublattice space and $\Pi_x$, $\Pi_y$, $\Pi_z$
in the valley space. Then, the first term in the Hamiltonian
may be written as ${\hat{h}}_{0} = - (1/2m) [ \sigma_x (p_x^2-p_y^2) +
2\Pi_z \sigma_y p_x p_y ]$. The operation of spatial inversion $i$
is represented by $\Pi_x \sigma_x$ because it swaps both valleys
and lattice sites, time inversion is given
by complex conjugation and $\Pi_x$, as it swaps valleys, too.
Hamiltonian~(\ref{heff1}) satisfies time-inversion symmetry
at zero magnetic field. The intrinsic terms
${\hat{h}}_{0}$, ${\hat{h}}_{w}$, ${\hat{h}}_{4}$, and ${\hat{h}}_{\Delta}$
satisfy spatial-inversion symmetry because the bilayer crystal structure
is spatial-inversion symmetric, but
terms ${\hat{h}}_{U}$ and ${\hat{h}}_{AB}$, with $\pm 1$ on the diagonal,
are imposed by external fields and they violate spatial-inversion symmetry, producing
a band gap between the conduction and valence bands.

\subsection{Interlayer coupling $\gamma_1$: massive chiral electrons}\label{s:mce}

The Hamiltonian ${\hat{h}}_{0}$ in equation~(\ref{heff1})
resembles the Dirac-like Hamiltonian of monolayer
graphene, but with a quadratic-in-momentum term on the off-diagonal rather than linear.
For example, the term $\pi^2/2m$ accounts for an effective hopping between
the non-dimer sites $A1$, $B2$ via the dimer sites $B1$, $A2$
consisting of a hop from $A1$ to $B1$ (contributing a factor $v\pi$),
followed by a transition between  $B1$, $A2$ dimer sites (giving a `mass' $\sim \gamma_1$),
and a hop from $A2$ to $B2$ (a second factor of $v\pi$).
The solutions are massive chiral electrons \cite{novo06,mcc06a}, with
parabolic dispersion $E = \pm p^2/2m$, $m = \gamma_1 /2v^2$.
The density of states is $m / (2\pi\hbar^2)$ per spin and per valley,
and the Fermi velocity $v_F = p_F/m$ is momentum dependent,
unlike the Fermi velocity $v$ of monolayer graphene.

The corresponding wave function is given by
\begin{eqnarray}
\psi &=& \frac{1}{\sqrt{2}} \left(
                                   \begin{array}{c}
                                     1 \\
                                     \mp e^{2i\xi\varphi} \\
                                   \end{array}
                                 \right) \, e^{i \mathbf{p}.\mathbf{r}/\hbar} \, . \label{wf0}
\end{eqnarray}
The wave function components describe the electronic amplitudes on the $A1$ and $B2$
sites, and it can be useful to introduce the concept of a pseudospin degree of
freedom \cite{novo06,mcc06a} that is related to these amplitudes.
If all the electronic density were located on the $A1$ sites, then the pseudospin
part of the wave function $| \!\!\uparrow \rangle = (1,0)$ could be viewed as a pseudospin `up' state, pointing
upwards out of the graphene plane.
Likewise, a state $| \!\!\downarrow \rangle = (0,1)$ with density solely on the $B2$ sites could be viewed
as a pseudospin `down' state.
However, density is usually shared equally between the two layers, so that the pseudospin
is a linear combination of up and down,
$ | \!\!\uparrow \rangle \mp e^{2i\xi\varphi} | \!\!\downarrow \rangle$, equation~(\ref{wf0}),
and it lies in the graphene plane.

The Hamiltonian may also be written as
${\hat{h}}_{0} = (p^2/2m) \,\mbox{\boldmath$\sigma$}.\mathbf{\hat n}_2$
where the pseudospin vector is
$\mbox{\boldmath$\sigma$} = \left( \sigma_x , \sigma_y , \sigma_z \right)$,
and $\mathbf{\hat n}_2 = - \left( \cos 2\varphi , \xi \sin 2\varphi , 0 \right)$
is a unit vector.
This illustrates the chiral nature of the electrons \cite{novo06,mcc06a}: the
chiral operator $\mbox{\boldmath$\sigma$}.\mathbf{\hat n}_2$ projects
the pseudospin onto the direction of quantization $\mathbf{\hat n}_2$,
which is fixed to lie in the graphene plane, but turns twice as quickly
as the momentum $\mathbf{p}$.
For these chiral quasiparticles, adiabatic propagation along a
closed orbit produces a Berry's phase \cite{berry84} change of $2\pi$ \cite{novo06,mcc06a}
of the wave function, in contrast to Berry phase $\pi$
in monolayer graphene.

Note that the chiral Hamiltonian ${\hat{h}}_{0}$
may be viewed as a generalisation of the Dirac-like Hamiltonian of monolayer graphene
and the second (after the monolayer) in a family of chiral Hamiltonians
$H_J$, $J = 1,2,\ldots$, corresponding to Berry's phase $J\pi$
which appear at low energy in ABC-stacked (rhombohedral) multilayer
graphene \cite{mcc06a,guinea06,koshino07,manes07,nak08,min08a,min08b,koshinoABC,zhang10}:
\begin{eqnarray}
H_J &=& g_J \left(
\begin{array}{cc}
0 & \left( {\pi }^{\dag }\right)^{J} \\
{\pi^{J}} & 0
\end{array} \right) ,
\end{eqnarray}
where $g_1 = v$ for monolayer, $g_2 = -1/2m$ for bilayer, and $g_3 = v^3/\gamma_1^2$
for trilayer graphene.
Since the pseudospin is related to the wavefunction amplitude on sites
that are located on different layers, pseudospin may be viewed as a
`which layer' degree of freedom \cite{min08a,zhang11}.

\subsection{Interlayer coupling $\gamma_3$: trigonal warping and the Lifshitz transition}\label{s:tw}

The Hamiltonian ${\hat{h}}_{0}$ in equation~(\ref{heff1})
yields a quadratic, isotropic dispersion relation $E = \pm p^2/2m$
with circular iso-energetic lines, {\em i.e.}, there is a circular Fermi line
around each $K$ point.
This is valid near the $K$ point, $p a / \hbar \ll 1$, whereas,
at high energy, and momentum $\mathbf{p}$ far from the $K$ point,
there is a triangular perturbation of the circular iso-energetic lines
known as trigonal warping, as in monolayer graphene and graphite.
It occurs because the band structure follows the symmetry of the crystal lattice
as described by the full momentum dependence
of the function $f ( \mathbf{k} )$, equation~(\ref{fk}) \cite{ando98}.

In bilayer graphene \cite{mcc06a}, as in bulk graphite \cite{dre74,nak76,ino62,gup72},
a second source of trigonal warping arises from
the skew interlayer coupling $\gamma_3$ between non-dimer $A1$ and $B2$ sites.
The influence of $\gamma_3$ on the band structure is described by equation~(\ref{efour}).
In the two-band Hamiltonian, it is described by ${\hat{h}}_{w}$ in equation~(\ref{heff1}),
the second term of which arises from a quadratic term in the expansion
of $f ( \mathbf{k} ) \approx - \sqrt{3}a( \xi p_x - i p_y)/2\hbar
+ a^2( \xi p_x + i p_y)^2/8\hbar^2$.
This second term has the same momentum dependence as ${\hat{h}}_{0}$, and, thus, it
actually only gives a small additional contribution to the mass $m$.
The first term in ${\hat{h}}_{w}$ causes trigonal warping of the iso-energetic lines
in directions $\varphi = \varphi_0$, where
$\varphi_0 = 0, {\textstyle\frac{2}{3}}\pi, {\textstyle\frac{4}{3}}\pi$ at $K_{+}$,
$\varphi_0 = {\textstyle\frac{1}{3}}\pi, \pi, {\textstyle\frac{5}{3}}\pi$ at $K_{-}$.

\begin{figure}[t]
\centerline{\epsfxsize=0.7\hsize \epsffile{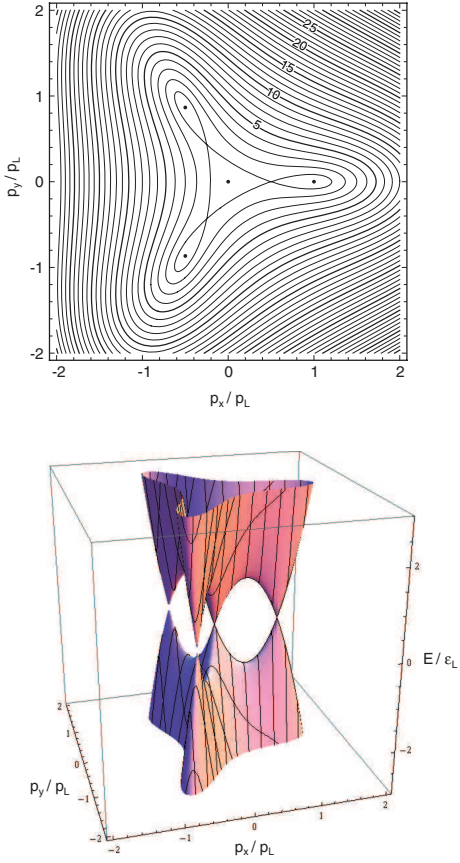}}
\caption{(a)
Trigonal warping of the equi-energy lines in the vicinity of each $K$ point,
and the Lifshitz transition in bilayer graphene. The energy is in units of $\vare_L$.
(b) Corresponding three-dimensional plot of the low-energy dispersion.}
\label{fig:trigonal}
\end{figure}

To analyse the influence of ${\hat{h}}_{w}$ at low energy, we
consider just ${\hat{h}}_{0}$ and the first term in ${\hat{h}}_{w}$, and
the resulting energy $E = \pm \varepsilon_1$ is given by
\begin{eqnarray}
\varepsilon_1^2 = \left( v_3 p \right)^2 - \frac{\xi v_3 p^3}{m} \cos \left( 3 \varphi \right)
+ \left( \frac{p^2}{2m} \right)^2 \, ,
\end{eqnarray}
in agreement with equation~(\ref{efour}) for $U=0$, $vp/\gamma_1 \ll 1$, and $v_3/v \ll 1$.
As it is linear in momentum, the influence of ${\hat{h}}_{w}$ and the resulting triangular
distortion of iso-energetic lines tend to increase as the
energy and momentum are decreased until a Lifshitz transition \cite{lif60} occurs at energy
\begin{equation}
 \varepsilon_L = \frac{\gamma_1}{4} \left(\frac{v_3}{v}\right)^2
\approx  1\,{\rm meV}.
\label{eq_lifshitz}
\end{equation}
For energies
$|E| < \varepsilon_L$, iso-energetic lines are broken into four separate
`pockets' consisting of one central pocket and three `leg' pockets, the latter centred
at momentum $p \approx \gamma_1 v_3/v^2$ and angle $\varphi_0$,
as shown in Figure~\ref{fig:trigonal}.
The central pocket is approximately circular for $|E| \ll \varepsilon_L$ with area
${\cal A}_c \approx \pi \varepsilon^2 / (\hbar v_3 )^2$, while each leg pocket
is approximately elliptical with area ${\cal A}_l \approx {\cal A}_c/3$.
Note that Berry phase $2\pi$ is conserved through the Lifshitz transition;
the three leg pockets each have Berry phase $\pi$
while the central pocket has $-\pi$ \cite{manes07,mik08}.

\subsection{Interlayer coupling $\gamma_4$ and on-site parameter $\Delta^{\prime}$:
electron-hole asymmetry}\label{s:gfour}

Skew interlayer coupling $\gamma_4$ between a non-dimer and a dimer site,
{\em i.e.}, between $A1$ and $A2$ sites or between $B1$ and $B2$ sites,
is described by ${\hat{h}}_{4}$ in equation~(\ref{heff1}),
where the effective velocity is $v_4 = \sqrt{3} a \gamma_4 / 2\hbar$.
This term produces electron-hole asymmetry in the band structure,
as illustrated by considering the energy eigenvalues
$E = \pm (p^2/2m) (1 \pm 2v_4/v)$ of the Hamiltonian
${\hat{h}}_{0} + {\hat{h}}_{4}$.
The energy difference $\Delta^{\prime}$ between dimer and non-dimer sites,
$\epsilon_{A1} = \epsilon_{B2} = 0$, $\epsilon_{B1} = \epsilon_{A2} = \Delta^{\prime}$,
equation~(\ref{defd}), also introduces electron-hole asymmetry into the
band structure: the low-energy bands described by ${\hat{h}}_{0} + {\hat{h}}_{\Delta}$
are given by $E = \pm p^2/2m (1 \pm \Delta^{\prime}/\gamma_1)$.

\subsection{Asymmetry between on-site energies: band gaps}\label{s:endif}

\subsubsection{Interlayer asymmetry}

Interlayer asymmetry $U$, equation~(\ref{defu}), describes a difference
in the on-site energies of the orbitals on the two layers
$\epsilon_{A1} = \epsilon_{B1} = - \epsilon_{A2} = - \epsilon_{B2} = -U/2$.
Its influence on the bands $E = \pm \varepsilon_{\alpha}(\mathbf{p})$
is described by equation~(\ref{efour}) with $v_3 = 0$:
\begin{eqnarray}
\varepsilon_{\alpha }^{2} = \frac{\gamma_{1}^{2}}{2}+\frac{{U}^{2}}{4}
+ v^{2}p^{2} + \left( -1\right)^{\alpha}
\sqrt{\frac{\gamma_{1}^{4}}{4}
+ v^{2}p^{2} \left[ \gamma_{1}^{2}+{U}^{2}\right]} \, , \nonumber \\
\label{efu}
\end{eqnarray}
The low-energy bands, $\alpha = 1$, display a distinctive `Mexican hat'
shape with a band gap $U_g$ between the conduction and valence bands
which occurs at momentum $p_g$ from the centre of the valley:
\begin{eqnarray}
U_g = \frac{|U| \gamma_1}{\sqrt{\gamma_1^2 + U^2}} \, ; \quad
p_g = \frac{|U|}{2v} \sqrt{\frac{2\gamma_1^2 + U^2}{\gamma_1^2 + U^2}} \, .
\end{eqnarray}
For small values of the interlayer asymmetry $U$, the band gap
is equal to the asymmetry $U_g = |U|$, but for large asymmetry values
$|U| \gg \gamma_1$ the band gap saturates $U_g \rightarrow \gamma_1$.
It is possible to induce interlayer asymmetry in bilayer graphene
through doping \cite{ohta06} or the use of external gates \cite{mcc06b,oostinga,castro}.
This is described in detail in section~\ref{s:tbg}.

\subsubsection{Intralayer asymmetry between $A$ and $B$ sites}

The energy difference $\delta_{AB}$ between $A$ and $B$ sites may be described
by the Hamiltonian~(\ref{Hbmin}) with
$\epsilon_{A1} = - \epsilon_{B1} = \epsilon_{A2} = - \epsilon_{B2} = \delta_{AB}/2$
and $v_3 = v_4 = 0$, yielding bands $E = \pm \varepsilon_\alpha$:
\begin{equation}
\varepsilon_{\alpha}^2 = \frac{\delta_{AB}^2}{4} + \frac{\gamma_1^2}{4}
\left[ \sqrt{1+4v^{2}p^{2}/\gamma _{1}^{2}} + \left(-1\right)^{\alpha} \right]^2 \, .
\end{equation}
Thus, $\delta_{AB}$ creates a band gap, but there is no Mexican hat structure.

\subsection{Next-nearest neighbour hopping}

The terms described in Hamiltonians~(\ref{Hbfull},\ref{Hbmin},\ref{heff1})
do not represent an exhaustive list of all possibilities. Additional coupling parameters
may be taken into account. For example, next-nearest neighbour hopping
within each layer \cite{wallace46,johnson,sasaki,peres06}
results in a term $(3 - |f(\mathbf{k})|^2)\gamma_n$ appearing
on every diagonal element of the Hamiltonian~(\ref{Hbfull}),
where $\gamma_n$ is the coupling parameter between next-nearest $A$ (or $B$ sites)
on each layer. Ignoring the constant-in-momentum part $3\gamma_n$
produces an additional term in the two-component Hamiltonian~(\ref{heff1})
\begin{eqnarray*}
{\hat{h}}_{n} &=& - \frac{\gamma_n v^2 p^2}{\gamma_0^2}
\left(
\begin{array}{cc}
1 & 0 \\
0 & 1
\end{array}
\right)  ,
\end{eqnarray*}
resulting in energies
$E = \pm p^2/2m (1 \mp \gamma_n\gamma_1/\gamma_0^2)$.
Thus, next-nearest neighbour hopping represents another source of
electron-hole asymmetry, after ${\hat{h}}_{4}$, ${\hat{h}}_{\Delta}$,
and $S_{\mathrm{b}}$.

\subsection{Spin-orbit coupling}

For monolayer graphene, Kane and Mele \cite{kanemele05} employed a symmetry
analysis to show that there are two distinct types of spin-orbit coupling
at the corners $K_{+}$ and $K_{-}$ of the Brillouin zone. These two types
of spin-orbit coupling exist in bilayer graphene, too.
In both monolayers and bilayers, the magnitude of spin-orbit
coupling - although the subject of theoretical debate - is generally
considered to be very small, with estimates roughly in the range of
$1$ to $100\,\mu$eV \cite{kanemele05,min06,dhh06,yao07,boett07,gmitra09,vgeld09,guinea10,liu10,mcc10,kon12}.

At the corner of the Brillouin zone $K_{\xi}$ in bilayer graphene,
the contribution of spin-orbit coupling to the two-component
low-energy Hamiltonian~(\ref{heff1}) may be written as
\begin{eqnarray}
{\hat{h}}_{SO} &=& \lambda_{SO} \xi \sigma_{z} S_{z} \, , \label{hso} \\
{\hat{h}}_{R} &=& \lambda_{R} \left( \xi \sigma_{x} S_{y} + \sigma_{y} S_{x} \right) \, , \label{hr}
\end{eqnarray}
where $\sigma_i$, $i = x,y,z$ are
Pauli spin matrices in the $A1$/$B2$ sublattice space,
and $S_j$, $j = x,y,z$ are Pauli spin matrices in the spin space.
The first term ${\hat{h}}_{SO}$ is intrinsic to graphene, \textit{i.e.}
it is a full invariant of the system.
Both intra- and inter-layer contributions to ${\hat{h}}_{SO}$ have been discussed
\cite{vgeld09,guinea10,liu10,kon12} with the dominant contribution to its magnitude $\lambda_{SO}$
attributed to skew interlayer coupling between $\pi$ and $\sigma$ orbitals \cite{guinea10,liu10}
or to the presence of unoccupied $d$ orbitals within each graphene layer \cite{kon12}.
Taken with the quadratic term ${\hat{h}}_{0}$ in the Hamiltonian~(\ref{heff1}),
${\hat{h}}_{SO}$ produces a gap of magnitude $2 \lambda_{SO}$ in the
spectrum of bilayer graphene,
but the two low-energy bands remain spin and valley degenerate (as in a monolayer):
$E = \pm \sqrt{\lambda_{SO}^2 + v^4p^4/\gamma_1^2}$.
However, there are gapless edge excitations and,
like monolayer graphene \cite{kanemele05},
bilayer graphene in the presence of intrinsic spin-orbit coupling is
a topological insulator with a finite spin Hall conductivity \cite{cort10,prada11}.

The second type of spin-orbit coupling is the Bychkov-Rashba term
${\hat{h}}_{R}$, equation~(\ref{hr}), which is permitted only if
mirror reflection symmetry with respect to the graphene plane is broken,
by the presence of an electric field or a substrate, say
\cite{rashba,kanemele05,min06,dhh06,rasbagraphene,rak10,vgeld09,cort10,prada11,qiao11b,mir12}.
Taken with the quadratic term ${\hat{h}}_{0}$ in the Hamiltonian~(\ref{heff1}),
${\hat{h}}_{R}$ does not produce a gap,
but, as in the monolayer, spin-splitting of magnitude $2 \lambda_{R}$ between the bands.
That is, there are four valley-degenerate bands at low energy,
\begin{eqnarray}
E^2 = \lambda_R^2 \left( \sqrt{1 + \frac{v^4p^4}{\lambda_R^2 \gamma_1^2}} \pm 1 \right)^2 \, .
\end{eqnarray}
Generally speaking, there is a rich interplay between tuneable interlayer
asymmetry $U$ and the influence of the intrinsic and the Bychkov-Rashba spin-orbit coupling
in bilayer graphene \cite{vgeld09,cort10,prada11,qiao11b,mir12}.
For example, the presence of interlayer
asymmetry $U$ breaks inversion symmetry and allows for spin-split levels in
the presence of intrinsic spin-orbit coupling only ($\lambda_R=0$) \cite{vgeld09},
while the combination of finite $U$ and very large Rashba coupling
has been predicted to lead to a topological insulator state even with
$\lambda_{SO}=0$ \cite{qiao11b}.

\subsection{The integer quantum Hall effect}
\label{s:qhe}

When a two-dimensional electron gas is placed in a perpendicular magnetic
field, electrons follow cyclotron orbits and their energies are quantised
as Landau levels \cite{landau}.
At a high enough magnetic field strength, the discrete nature of the
Landau level spectrum is manifest as the integer quantum Hall
effect \cite{vk80,p+r87,macdonald89}, whereby the Hall conductivity assumes
values that are integer multiples of the quantum of conductivity $e^2/h$.

The Landau level spectrum of monolayer graphene was calculated by
McClure \cite{mcclure56} nearly sixty years ago,
and there have been a number of related theoretical studies
\cite{semenoff84,haldane88,zheng02,gusynin05,peres06,herbut07}
considering the consequences of chirality in graphene.
The experimental observation of the integer quantum Hall effect in monolayer
graphene \cite{novo05,zhang05} found an unusual sequencing of the
quantised plateaus of Hall conductivity, confirming the chiral nature
of the electrons and prompting an explosion of interest in the field \cite{cnreview}.
In bilayer graphene, the observation of the integer quantum Hall
effect \cite{novo06} and the calculated Landau level spectrum  \cite{mcc06a}
uncovered additional features related to the chiral nature of the electrons.

\subsubsection{The Landau level spectrum of bilayer graphene}
\label{ss:llbi}

We consider the Landau level spectrum of the two-component chiral
Hamiltonian ${\hat{h}}_{0}$, equation~(\ref{heff1}).
The magnetic field is accounted for by the operator
$\mathbf{p} = (p_x , p_y) \equiv -i\hbar \nabla + e \mathbf{A}$
where $\mathbf{A}$ is the vector potential and the charge of the electron is $- e$.
For a magnetic field perpendicular to the bilayer,
$\mathbf{B} = \left( 0 , 0, - B \right)$ where $B = |\mathbf{B}|$,
the vector potential may be written in the
Landau gauge $\mathbf{A} = \left( 0, -Bx, 0 \right)$,
which preserves translational invariance in the $y$ direction.
Then,
$\pi = -i \hbar \xi \partial_x + \hbar \partial_y - i e B x$
and
$\pi^{\dagger} = -i \hbar \xi \partial_x - \hbar \partial_y + i e B x$,
and eigenstates are comprised of functions that are harmonic oscillator states
in the $x$ direction and plane waves in the $y$ direction
\cite{p+r87,macdonald89},
\begin{eqnarray}
\phi_{\ell} \left( x , y \right) &=&
A_{\ell} \, {\cal H}_{\ell} \!\! \left( \frac{x}{\lambda_B} - \frac{p_y \lambda_B}{\hbar} \right) \nonumber \\
&& \times
\exp \left[ - \frac{1}{2}\left( \frac{x}{\lambda_B} - \frac{p_y \lambda_B}{\hbar} \right)^2
+ i \frac{p_y y}{\hbar} \right] \!\! , \label{mcc:ho}
\end{eqnarray}
where the magnetic length is $\lambda_B = \sqrt{\hbar/eB}$,
${\cal H}_{\ell}$ are Hermite polynomials of order ${\ell}$ for integer
${\ell} \geq 0$, and
$A_{\ell} = 1/\sqrt{2^{\ell} {\ell}! \sqrt{\pi}}$ is a normalisation constant.

The operators $\pi$ and $\pi^{\dagger}$ appearing in the Hamiltonian~(\ref{heff1})
act as raising and lowering operators for the harmonic oscillator states~(\ref{mcc:ho}).
At the first valley, $K_{+}$,
\begin{eqnarray}
K_{+}: \qquad \,\,\, \pi \phi_{\ell} &=& - \frac{\sqrt{2} i \hbar}{\lambda_B} \sqrt{{\ell}} \, \phi_{{\ell}-1} \, ,
\label{mcc:lower} \\
K_{+}: \qquad \pi^{\dagger} \phi_{\ell} &=& \frac{\sqrt{2} i \hbar}{\lambda_B} \sqrt{{\ell}+1} \, \phi_{{\ell}+1} \, ,
\label{mcc:raise}
\end{eqnarray}
and $\pi \phi_{0}=0$.
Then, it is possible to show that the Landau level spectrum of the Hamiltonian~(\ref{heff1})
consists of a series of electron and hole levels with energies and wave functions \cite{mcc06a} given
by
\begin{eqnarray}
E_{{\ell},\pm} &=&
\pm \hbar \omega_c \sqrt{{\ell} ({\ell}-1)}  \, , \quad {\ell} \geq 2 , \label{mcc:bikp} \\
K_{+}: \qquad \psi_{{\ell},\pm} &=& \frac{1}{\sqrt{2}} \left(
                                  \begin{array}{c}
                                    \phi_{\ell} \\
                                    \pm \phi_{{\ell}-2} \\
                                  \end{array}
                                \right)  ,  \quad {\ell} \geq 2 , \label{mcc:stateskp}
\end{eqnarray}
where $\omega_c = eB/m$
and $\pm$ refer to the electron and hole states, respectively.
For high values of the index, ${\ell} \gg 1$,
the levels are approximately equidistant
with spacing $\hbar \omega_c$ proportional to the magnetic
field strength $B$.
However, this spectrum, equation~(\ref{mcc:bikp}),
is only valid for sufficiently
small level index and magnetic field ${\ell} \hbar \omega_c \ll \gamma_1$
because the effective Hamiltonian~(\ref{heff1}) is only applicable at low energy.

In addition to the field-dependent levels, there are two levels
fixed at zero energy $E_1 = E_0 = 0$ with eigenfunctions:
\begin{eqnarray}
K_{+}: \qquad \psi_{1} = \left(
                                  \begin{array}{c}
                                    \phi_1 \\
                                    0 \\
                                  \end{array}
                                \right) , \qquad
\psi_{0} = \left(
                                  \begin{array}{c}
                                    \phi_0 \\
                                    0 \\
                                  \end{array}
                                \right) , \label{mcc:bikp10}
\end{eqnarray}
They may be viewed as arising from the square of
the lowering operator in the Hamiltonian~(\ref{heff1})
which acts on both the oscillator ground state
and the first excited state to give zero energy
$\pi^2 \phi_{0} = \pi^2 \phi_{1} = 0$.
The eigenfunctions $\psi_0$ and $\psi_1$
have a finite amplitude on the $A1$ sublattice, on the bottom
layer, but zero amplitude on the $B2$ sublattice.

At the second valley, $K_{-}$, the roles of operators
$\pi$ and $\pi^{\dagger}$ are reversed:
\begin{eqnarray}
K_{-}: \qquad \,\,\, \pi \phi_{\ell} &=& - \frac{\sqrt{2} i \hbar}{\lambda_B} \sqrt{{\ell}+1} \, \phi_{{\ell}+1} \, ,
\label{mcc:lowerm} \\
K_{-}: \qquad \pi^{\dagger} \phi_{\ell} &=& \frac{\sqrt{2} i \hbar}{\lambda_B} \sqrt{{\ell}} \, \phi_{{\ell}-1} \, ,
\label{mcc:raisem}
\end{eqnarray}
and $\pi^{\dagger} \phi_{0}=0$.
The Landau level spectrum at $K_{-}$ is degenerate
with that at $K_{+}$, \textit{i.e.}, $E_{{\ell},\pm} = \pm \hbar \omega_c \sqrt{{\ell} ({\ell}-1)}$ for
${\ell} \geq 2$ and $E_1 = E_0 = 0$,
but the roles of the $A1$ and $B2$ sublattices are reversed:
\begin{eqnarray}
K_{-}: \qquad \psi_{{\ell},\pm} &=& \frac{1}{\sqrt{2}} \left(
                                  \begin{array}{c}
                                    \phi_{\ell-2} \\
                                    \pm \phi_{{\ell}} \\
                                  \end{array}
                                \right)  ,  \quad {\ell} \geq 2 , \label{mcc:stateskm} \\
K_{-}: \qquad \,\,\,\,\, \psi_{1} &=& \left(
                                  \begin{array}{c}
                                    0 \\
                                    \phi_1 \\
                                  \end{array}
                                \right) , \qquad
\psi_{0} = \left(
                                  \begin{array}{c}
                                    0 \\
                                    \phi_0 \\
                                  \end{array}
                                \right) , \label{mcc:bikm10}
\end{eqnarray}
The valley structure and electronic spin each contribute
a twofold degeneracy to the Landau level spectrum.
Thus, each level in bilayer level graphene is fourfold degenerate,
except for the zero energy levels which have eightfold degeneracy
due to valley, spin and the orbital degeneracy of $\psi_0$, $\psi_1$.

\subsubsection{Three types of integer quantum Hall effect}
\label{ss:qhebi}

The form of the Landau level spectrum is manifest in a measurement
of the integer quantum Hall effect.
Here, we will compare the density dependence of the Hall conductivity
$\sigma_{xy}(n)$ for bilayer graphene with that of a conventional semiconductor
and of monolayer graphene.

The Landau level spectrum of a conventional two-dimensional semiconductor
is $E_{\ell} = \hbar \omega_c ({\ell} + 1/2)$, $\ell \geq 0$,
where $\omega_c = eB/m$ is the cyclotron frequency \cite{p+r87,macdonald89}.
As density is changed, there is a step in $\sigma_{xy}$ whenever
a Landau level is crossed, and the separation of steps on
the density axis is equal to the maximum carrier density per Landau level,
$g B / \varphi_0$, where $\varphi_0 = h/e$ is the flux quantum
and $g$ is a degeneracy factor.
Each plateau of the Hall conductivity $\sigma_{xy}$ occurs at a quantised value
of $N\!ge^2/h$ where $N$ is an integer labelling the plateau and
$g$ is an integer describing the level degeneracy;
steps between adjacent plateaus have height $ge^2/h$.

\begin{figure}[t]
\centerline{\epsfxsize=0.9\hsize \epsffile{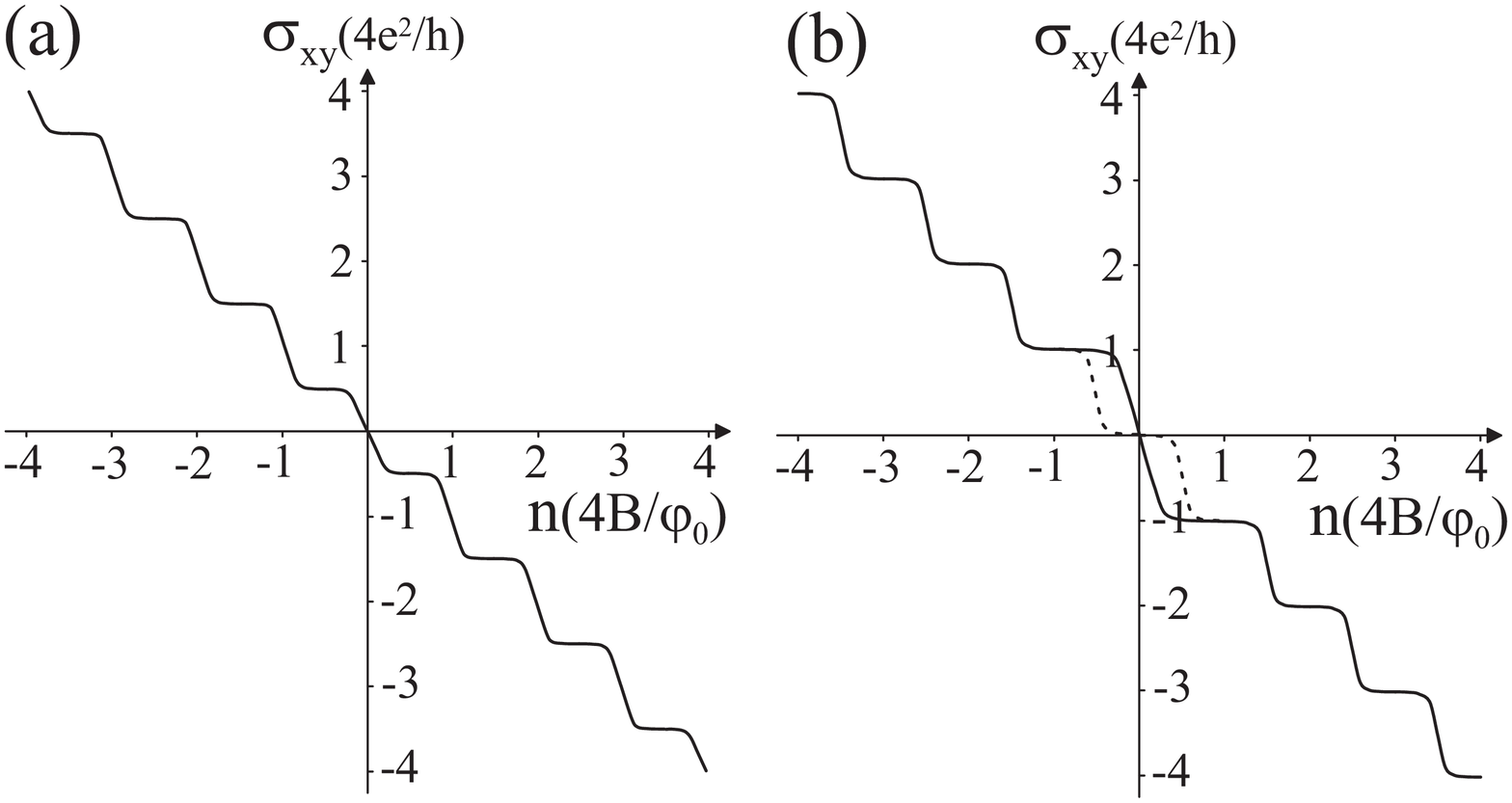}}
\caption{Schematic of the dependence of the
Hall conductivity $\sigma_{xy}$ on carrier density $n$ for
(a) monolayer graphene and (b) bilayer graphene,
where $\varphi_0 = h/e$ is the flux quantum and $B$ is the magnetic field strength.
The dashed line in (b) shows the behaviour for a conventional semiconductor
or bilayer graphene in the presence of large interlayer asymmetry $U$
(section~\ref{ss:ia}) with fourfold level degeneracy due to spin and valley degrees of freedom.}
\label{fig:qhe}
\end{figure}

The Landau level spectrum of monolayer
graphene \cite{mcclure56,zheng02,gusynin05,peres06,herbut07} consists of
an electron and a hole series of levels, $E_{{\ell},\pm} = \pm \sqrt{2{\ell}} \hbar v/\lambda_B$
for ${\ell} \geq 1$, with an additional level at zero energy $E_{0} = 0$.
All of the levels are fourfold degenerate, due to spin and valley degrees of freedom.
The corresponding Hall conductivity is shown schematically in figure~\ref{fig:qhe}(a).
There are steps of height $4e^2/h$ between each plateau, as expected by consideration of
the conventional case, but the plateaus occur at half-integer values of
$4e^2/h$ instead of integer ones, as observed experimentally \cite{novo05,zhang05}.
This is due to the existence of the fourfold-degenerate level $E_{0} = 0$ at
zero energy, which contributes to a step of height $4e^2/h$ at zero density.

For bilayer graphene, plateaus in the Hall conductivity $\sigma_{xy} (n)$,
Fig.~\ref{fig:qhe}(b), occur at integer multiples of $4e^2/h$.
This is similar to a conventional semiconductor with level degeneracy
$g=4$ arising from the spin and valley degrees of freedom.
Deviation from the conventional case occurs at low density. In the bilayer
there is a step in $\sigma_{xy}$ of height $8e^2/h$
across zero density, accompanied by a plateau separation of $8B/\varphi_0$
in density \cite{novo06,mcc06a}, arising from the
eightfold degeneracy of the zero-energy Landau levels. This is shown as
the solid line in Fig.~\ref{fig:qhe}(b), whereas, for a conventional semiconductor,
there no step across zero density (the dashed line).

Thus, the chirality of charge carriers in monolayer and bilayer graphene
give rise to four- and eight-fold degenerate Landau levels
at zero energy and to steps of height of four and eight times
the conductance quantum $e^2/h$ in the Hall conductivity at zero
density \cite{novo05,zhang05,novo06}.
Here, we have assumed that the degeneracy of the Landau levels
is preserved, \textit{i.e.}, any splitting of the levels is
negligible as compared to temperature and level broadening in an experiment.
The role of electronic interactions in bilayer graphene is described in section~\ref{s:eei},
while we discuss the influence of interlayer asymmetry on the Landau level
spectrum and integer quantum Hall effect in the next section.

\subsubsection{The role of interlayer asymmetry}
\label{ss:ia}

The Landau level states, equations~(\ref{mcc:stateskp},\ref{mcc:stateskm}),
have different amplitudes on the lower ($A1$ sublattice) and
upper ($B2$ sublattice) layers, with the role of the sublattice sites
swapped at the two valleys. Thus, interlayer asymmetry $U$
as described by the effective Hamiltonian ${\hat{h}}_{U}$, equation~(\ref{heff1}),
leads to a weak splitting of the valley degeneracy of the levels \cite{mcc06a}:
\begin{eqnarray}
E_{{\ell},\pm} \approx \pm \hbar \omega_c \sqrt{{\ell} ({\ell}-1)}
+ \frac{\xi U \hbar \omega_c}{2\gamma_1} \, .
\label{mcc:stateshigh}
\end{eqnarray}
Such splitting is prominent for the zero-energy states \cite{mcc06a},
equations~(\ref{mcc:bikp10},\ref{mcc:bikm10}), because they
only have non-zero amplitude on one of the layers, depending on the valley:
\begin{eqnarray}
E_0 &=& - \frac{1}{2} \xi U \, , \label{mcc:stateslow} \\
E_1 &=& - \frac{1}{2} \xi U + \frac{\xi U \hbar \omega_c}{\gamma_1} \, .
\end{eqnarray}
When the asymmetry is large enough,
then the splitting $U$ of the zero energy levels from each valley
results in a sequence of quantum Hall plateaus
at all integer values of $4e^2 /h$ including a plateau at zero density \cite{mcc06b},
as observed experimentally \cite{castro}.
This behaviour is shown schematically as the dashed line in figure~\ref{fig:qhe}(b).
The Landau level spectrum in the presence of large interlayer asymmetry $U$
has been calculated \cite{mp07,mucha09,nakamura-castro09,zhang-fogler11},
including an analysis of level crossings \cite{zhang-fogler11}
and a self-consistent calculation of the
spectrum in the presence of an external gate \cite{mucha09,nakamura-castro09},
generalising the zero-field procedure outlined in the next section.

\section{Tuneable band gap}\label{s:tbg}

\subsection{Experiments}

A tuneable band gap in bilayer graphene was first observed with angle-resolved
photoemission of epitaxial bilayer graphene on silicon carbide \cite{ohta06},
and the ability to control the gap was demonstrated by doping with potassium.
Since then, the majority of experiments probing the band gap have used
single or dual gate devices based on exfoliated bilayer graphene flakes \cite{oostinga,castro}.
The band gap has now been observed in a number of different
experiments including photoemission \cite{ohta06},
magnetotransport \cite{castro},
infrared spectroscopy \cite{hen08,zhang08,kuz09a,mak09,zhang09,kuz09b},
electronic compressibility \cite{hen10,young10},
scanning tunnelling spectroscopy \cite{desh09},
and transport \cite{oostinga,xia10,sza10,zou10,miyazaki10,jing10,tay10,yan10}.

The gap observed in optics \cite{ohta06,hen08,zhang08,kuz09a,mak09,zhang09,kuz09b}
is up to $250\,$meV \cite{mak09,zhang09}, the value expected theoretically (as the
gap should saturate at the value of the interlayer coupling $\gamma_1$).
Transport measurements show insulating
behaviour \cite{oostinga,xia10,sza10,zou10,miyazaki10,jing10,tay10,yan10}, but,
generally, not the huge suppression of conductivity expected for a gap of this magnitude,
and this has been attributed to edge states \cite{li-martin11},
the presence of disorder \cite{castro10c,rossi11,abergel12} or disorder and chiral charge carriers \cite{trushin12}.
Broadly speaking, transport seems to occur through different mechanisms in different
temperature regimes with thermal activation \cite{xia10,sza10,zou10,miyazaki10,tay10,yan10}
at high temperature (above, roughly, $2$ to $50\,$K)
and variable-range \cite{oostinga,zou10,miyazaki10,jing10,yan10}
or nearest-neighbour hopping \cite{zou10,tay10} at
low temperature.

\subsection{Hartree model of screening}

External gates are generally used to control the density of electrons $n$
on a graphene device \cite{novo04}, but, in bilayer graphene,
external gates will also place the separate layers of the bilayer
at different potential energies resulting in
interlayer asymmetry $U = \epsilon_{2} - \epsilon_{1}$
(where $\epsilon_{1} = \epsilon_{A1} = \epsilon_{B1} = -U/2$
and $\epsilon_{2} = \epsilon_{A2} = \epsilon_{B2} = U/2$).
Thus, changing the applied gate voltage(s) will tend to tune both the density $n$
and the interlayer asymmetry $U$, and, ultimately,
the band gap $U_g$. The dependence of the band gap on
the density $U_g (n)$ relies on screening by electrons on the bilayer.
In the following, we describe a simple model
\cite{mcc06b,min07,mcc07,mcc07b,castro,zhang08,fogler10,castro10}
that has been developed to take into account screening
using the tight-binding model and Hartree theory.

\subsubsection{Electrostatics: asymmetry parameter, layer densities and external gates}

\begin{figure}[t]
\centerline{\epsfxsize=1.0\hsize \epsffile{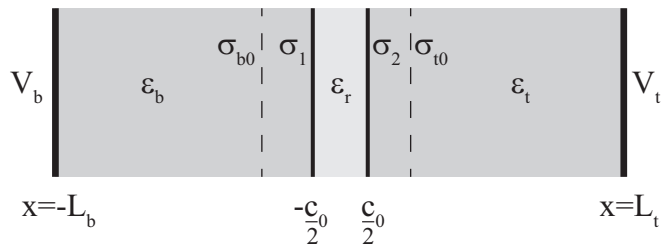}}
\caption{Schematic of bilayer graphene in the presence of external gates
located at $x=-L_b$ and $x= +L_t$, with potentials $V_b$ and $V_t$, which
are separated from the bilayer by media of dielectric constants
$\varepsilon_b$ and $\varepsilon_t$, respectively. The bilayer is modelled
as two parallel conducting plates positioned at $x=-c_0/2$ and $+c_0/2$,
separated by a region of dielectric constant $\varepsilon_r$.
The layers have charge densities $\sigma_1 = -e n_1$ and $\sigma_2 = -e n_2$
corresponding to layer number densities $n_1$ and $n_2$.
Charge densities $\sigma_{b0}$ and $\sigma_{t0}$ (dashed lines) arise from the
presence of additional charged impurities.}
\label{fig:gates}
\end{figure}

We use the SI system of units, and the electronic charge is $-e$ where $e > 0$.
The bilayer graphene device is modelled as two parallel conducting plates
that are very narrow in the $x$-direction, continuous and infinite in
the $y$-$z$ plane, positioned at $x=-c_0/2$ and $+c_0/2$,
figure~\ref{fig:gates}.
Here, $c_0$ is the interlayer spacing and we denote the dielectric constant
of the interlayer space as $\varepsilon_r$ (it doesn't include the
screening by $\pi$-band electrons that we are explicitly modelling).
Layer number densities are $n_1$ and $n_2$, with corresponding charge densities
$\sigma_1 = -e n_1$ and $\sigma_2 = -e n_2$.

We take into account the presence of a back gate and a top gate, infinite in the $y$-$z$ plane,
located at $x=-L_b$ and $x= +L_t$, with potentials $V_b$ and $V_t$, which
are separated from the bilayer by media of dielectric constants
$\varepsilon_b$ and $\varepsilon_t$, respectively.
It is possible to describe the presence of additional charge near
the bilayer - due to impurities, say - by introducing density $n_{b0}$
on the back-gate side and $n_{t0}$ on the top-gate side.
They correspond to charge densities $\sigma_{b0} = e n_{b0}$ and $\sigma_{t0} = e n_{t0}$,
assuming that $n_{b0}$ and $n_{t0}$ are positive for positive charge.

Using Gauss's Law, it is possible to relate the external gate potentials
and impurities concentrations to the layer densities and the interlayer
asymmetry \cite{mcc06b,kosh09a,fogler10,mcc12}:
\begin{eqnarray}
n = n_1 + n_2
&=& \frac{\varepsilon_0 \varepsilon_b V_b}{eL_b}
+ \frac{\varepsilon_0 \varepsilon_t V_t}{eL_t}
+ n_{b0} + n_{t0} \, , \label{VA} \\
U
&=& - \frac{\varepsilon_t}{\varepsilon_r}\frac{c_0}{L_t}e V_t + \frac{e^2 c_0}{\varepsilon_0 \varepsilon_r} \left( n_2 - n_{t0} \right) \, , \label{VB}
\end{eqnarray}
where the field within the bilayer interlayer space is approximately
equal to $U / ( e c_0 )$.
Equation~(\ref{VA}) expresses the total electron density
$n = n_1 + n_2$ in terms of the external potentials, generalising
the relation for monolayer graphene \cite{novo04}.
Note that the background densities $n_{b0}$ and $n_{t0}$ shift the effective values of
the gate potentials $V_b$ and $V_t$.
The second equation, for $U$, may be rewritten as
\begin{eqnarray}
U &=& U_{\mathrm{ext}}
+ \Lambda \gamma_1 \frac{\left( n_2 - n_1 \right)}{n_{\perp}} \, , \label{VC} \\
U_{\mathrm{ext}}
&=& \frac{e c_0}{2 \varepsilon_r} \! \left(
\frac{\varepsilon_b}{L_b} V_b
- \frac{\varepsilon_t}{L_t} V_t \right)
+ \Lambda \gamma_1 \frac{\left( n_{b0} - n_{t0} \right)}{n_{\perp}} , \label{VD}
\end{eqnarray}
where the characteristic density scale $n_{\perp}$ and the
dimensionless screening parameter $\Lambda$ are
\begin{eqnarray}
n_{\perp} = \frac{\gamma_1^2}{\pi \hbar^2v^2} \, , \qquad
\Lambda = \frac{c_0 e^2 \gamma_1}{2\pi \hbar^2v^2 \varepsilon_0 \varepsilon_r}
\equiv \frac{c_0 e^2 n_{\perp}}{2 \gamma_1 \varepsilon_0 \varepsilon_r} \, .
\end{eqnarray}
Equation~(\ref{VC}) relates the asymmetry parameter $U$ to a sum of its
value, $U_{\mathrm{ext}}$, if screening were negligible
plus a term accounting for screening.
Parameter values $\gamma_1 = 0.39$eV and $v = 1.0 \times 10^6$ms$^{-1}$
give $n_{\perp} = 1.1 \times 10^{13}$cm$^{-2}$.
With interlayer spacing $c_0 = 3.35${\AA} and an estimate
for the dielectric constant of $\varepsilon_r \approx 1$,
then $\Lambda \sim 1$, showing that screening is relevant.

\subsubsection{Calculation of individual layer densities}

Through electrostatics, the asymmetry parameter $U$ is related to
layer densities $n_{1}$ and $n_{2}$, equation~(\ref{VC}).
The densities $n_{1}$ and $n_{2}$ also depend on $U$ because of
the band structure of bilayer graphene.
Analytical calculations are possible \cite{mcc06b,min07,fogler10}
if only the dominant inter-layer coupling $\gamma_1$
is taken into account in the four-band Hamiltonian~(\ref{Hbmin}).
Here we will use the two-band model~(\ref{heff1})
with an explicit ultraviolet cutoff \cite{min07} when integrating over the whole Brillouin
zone.
The simplified two-component Hamiltonian is
\begin{eqnarray}
{\hat{H}}_{2} \approx -\frac{v^2}{\gamma_1}\left(
\begin{array}{cc}
0 & \left( {\pi }^{\dag }\right) ^{2} \\
{\pi ^{2}} & 0
\end{array} \right)
- \frac{U}{2} \left(
                \begin{array}{cc}
                  1 & 0 \\
                  0 & -1 \\
                \end{array}
              \right) \, , \label{hsim}
\end{eqnarray}
Solutions to the energy eigenvalue equation ${\hat{H}}_{2} \psi = E \psi$
are given by
\begin{eqnarray}
E &=& \pm \sqrt{ \frac{U^2}{4} + \frac{v^4p^4}{\gamma_1^2} } \, , \\
\psi &=& \sqrt{\frac{E - U/2}{2E}} \left(
                                   \begin{array}{c}
                                     1 \\
                                     - \frac{v^2p^2}{\gamma_1 \left( E-U/2 \right)}e^{2i\xi\varphi} \\
                                   \end{array}
                                 \right) \, e^{i \mathbf{p}.\mathbf{r}/\hbar} \, . \label{wf}
\end{eqnarray}
Layer densities are determined by integration over the
circular Fermi surface
\begin{eqnarray}
n_{1(2)} = \frac{2}{\pi \hbar^2} \int | \psi_{A1(B2)} ( p ) |^2 p \, dp \, ,
\end{eqnarray}
where a factor of four takes
account of spin and valley degeneracy.

For simplicity, we assume the Fermi energy lies within the conduction band.
Using the solution~(\ref{wf}), the contribution of the
partially-filled conduction band to the
layer densities \cite{mcc06b,falk09,fogler10} is given by
\begin{eqnarray}
n_{1(2)}^{\mathrm{cb}} &=&  \frac{1}{\pi \hbar^2} \int p d p
\left( \frac{E \mp U/2}{E} \right) , \label{nlayercb} \\
&=& \frac{1}{\pi\hbar^2} \int_{0}^{p_F} p d p
\mp \frac{U}{2\pi\hbar^2} \int_{0}^{p_F} \frac{ p d p}{\sqrt{U^2/4 + v^4p^4/\gamma_1^2}} \, , \nonumber \\
&\approx& \frac{n}{2}
\mp \frac{n_{\perp} U}{4\gamma_1} \ln
\left( \frac{2 |n| \gamma_1}{n_{\perp}|U|}
+ \sqrt{1 + \left( \frac{2 n\gamma_1}{n_{\perp} U} \right)^2} \right)  , \nonumber
\end{eqnarray}
where the minus (plus) sign is for the first (second) layer,
$p_F$ is the Fermi momentum,
and the total density is $n = p_F^2 / \pi \hbar^2$ measured with
respect to the charge neutrality point, {\em i.e.}, we assume
that the point of zero density is realised when the Fermi level
lies at the crossing point of the conduction and valence bands.

In addition, we take into account the contribution of the filled
valence band $n_{1(2)}^{\mathrm{vb}}$ to the individual layer densities.
Note that, as the asymmetry parameter $U$ varies, the filled valence band
does not contribute to any change in the total density $n$, but it does
contribute to the difference $n_{1} - n_{2}$.
This may be obtained by integrating with respect to momentum
as in equation~(\ref{nlayercb}), but introducing an ultraviolet
cutoff $p_{\mathrm{max}} = \gamma_1/v$ equivalent to
$n_{\mathrm{max}} = n_{\perp}$ \cite{min07}. Then,
the contribution of the filled valence band \cite{mcc06b,min07,mcc07,mcc07b,fogler10,castro10}
is given by
\begin{eqnarray}
n_{1(2)}^{\mathrm{vb}} &\approx&
\pm \frac{U}{2\pi\hbar^2} \int_{0}^{p_{\mathrm{max}}} \frac{ p d p}{\sqrt{U^2/4 + v^4p^4/\gamma_1^2}} \, , \nonumber \\
&\approx& \pm \frac{n_{\perp}U}{4\gamma_1} \ln \left( \frac{4\gamma_1}{|U|} \right) \, . \label{nlayervb}
\end{eqnarray}
Combining the contributions of equations~(\ref{nlayercb}) and~(\ref{nlayervb}), the
individual layer density,
$n_{1(2)} = n_{1(2)}^{\mathrm{cb}} + n_{1(2)}^{\mathrm{vb}}$, is given \cite{mcc06b,fogler10} by
\begin{eqnarray}
\!\!\!\!\!\!\!\!\!n_{1(2)} \!\approx\! \frac{n}{2}
\!\mp\! \frac{n_{\perp} U}{4\gamma_1} \ln \!\!
\left( \! \! \frac{|n|}{2n_{\perp}} \!+\!
\frac{1}{2}\sqrt{ \left( \frac{n}{n_{\perp}} \right)^2 + \left( \frac{U}{2\gamma_1} \right)^2} \right)\!\! . \label{n12}
\end{eqnarray}
Note that some calculations \cite{falk09} include only the contribution~(\ref{nlayercb})
of the partially-filled conduction band,
others \cite{min07} include only the filled valence band~(\ref{nlayervb}).

\subsubsection{Self-consistent screening}

\begin{figure}[t]
\centerline{\epsfxsize=0.8\hsize \epsffile{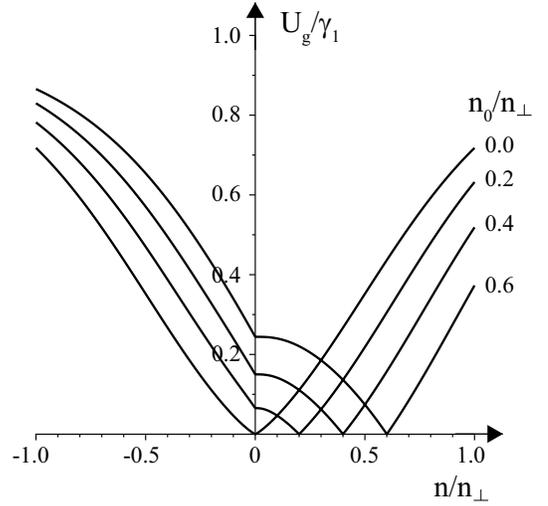}}
\caption{Density-dependence of the band gap $U_g (n)$ in bilayer graphene
as the back-gate voltage $V_b$ is varied for a fixed top-gate voltage $V_t$ \cite{mcc06b}.
The effective offset-density is
$n_0 = 2[\varepsilon_0 \varepsilon_t V_t/(eL_t) + n_{t0}]$.
Plots were made for screening parameter $\Lambda = 1$,
using $U_g = |U|\gamma_1/\sqrt{U^2 + \gamma_1^2}$
and numerical solution of equation~(\ref{delta}).}
\label{fig:gateplot}
\end{figure}

\begin{figure}[t]
\centerline{\epsfxsize=0.9\hsize \epsffile{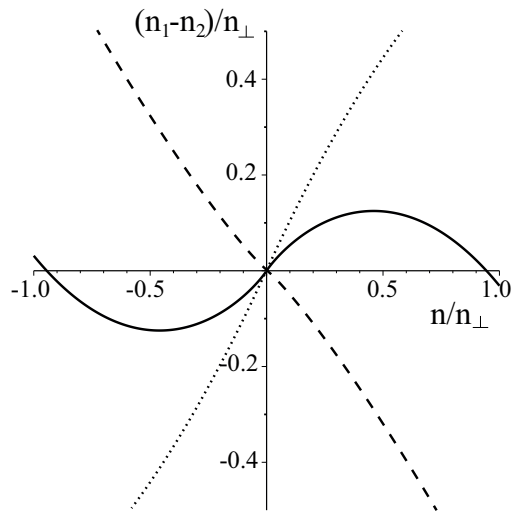}}
\caption{Density-dependence of the difference in layer
densities $n_1 - n_2$ in bilayer graphene
as the back-gate voltage $V_b$ is varied for a fixed top-gate voltage $V_t$ \cite{mcc06b}.
Plots were made for screening parameter $\Lambda = 1$
and the effective offset-density is
$n_0 = 2[\varepsilon_0 \varepsilon_t V_t/(eL_t) + n_{t0}] = 0$,
corresponding to the data labelled $n_0 = 0$ in Figure~\ref{fig:gateplot},
obtained by numerical solution of equation~(\ref{delta}).
The dashed line shows the contribution of the partially filled bands as measured with respect
to the charge neutrality point~~(\ref{nlayercb}) \cite{falk09}, the dotted line shows the contribution
of the full valence band~(\ref{nlayervb}) \cite{min07}, the solid line is their sum~(\ref{n12}) \cite{mcc06b,fogler10}.
}
\label{fig:densities}
\end{figure}

Substituting the layer density~(\ref{n12}) into the expression~(\ref{VC})
describing screening gives an expression \cite{mcc06b,fogler10,mcc12} for the
density-dependence of the asymmetry parameter $U$:
\begin{eqnarray}
\frac{U(n)}{U_{\mathrm{ext}}} \approx \left[
1 - \frac{\Lambda}{2}
\ln
\left( \frac{|n|}{2n_{\perp}} +
\frac{1}{2}\sqrt{ \left( \frac{n}{n_{\perp}} \right)^2 + \left( \frac{U}{2\gamma_1} \right)^2} \right) \!
\right]^{-1} \!\!\! , \nonumber \\ \label{delta}
\end{eqnarray}
where $U_{\mathrm{ext}}$ is the asymmetry in the absence of
screening~(\ref{VD}).
The extent of screening is described by the logarithmic
term with argument depending on $n$ and $U$,
and a prefactor proportional to the screening parameter
$\Lambda \sim 1$ (as discussed earlier).
A common experimental setup, especially for exfoliated graphene on a silicon substrate
\cite{novo04,novo05,zhang05,novo06}, includes a single back gate.
Figure~\ref{fig:gateplot} shows the density-dependence of the band gap $U_g (n)$
plotted as the back-gate voltage $V_b$ is varied for a fixed top-gate voltage $V_t$.
In this case, the influence of the top-gate voltage $V_t$ may be absorbed
into an effective offset-density
$n_0 = 2[\varepsilon_0 \varepsilon_t V_t/(eL_t) + n_{t0}]$ \cite{mcc06b}
giving $U_{\mathrm{ext}} = \Lambda \gamma_1 (n - n_0)/n_{\perp}$ in
equation~\ref{delta}. Figure~\ref{fig:densities} shows the dependence
of the difference in layer densities $n_1 - n_2$ for the case $n_0 = 0$
including both the contribution of the partially filled bands as measured with respect
to the charge neutrality point~(\ref{nlayercb}) (dashed line)
and the contribution of the full valence band~(\ref{nlayervb}) (dotted line).
The sum of both terms (solid line) shows that $n_1 - n_2$ is positive (negative)
for positive (negative) total density $n$. Recalling that layer $1$ is closest to
the back gate, this shows that the bilayer is polarised along the electric field,
as expected \cite{fogler10}.

A single back gate in the absence of additional charged dopants
may be described by $V_t = n_{b0} = n_{t0} = 0$,
resulting in simplified expressions
$n = \varepsilon_0 \varepsilon_b V_b/(eL_b)$
(as in monolayer graphene \cite{novo04})
and $U_{\mathrm{ext}} = \Lambda \gamma_1 n / n_{\perp}$.
Using $|U| \ll \gamma_1$, equation~(\ref{delta})
simplifies \cite{mcc06b,mcc07,mcc07b,mcc12} as
\begin{eqnarray}
U (n) \approx \frac{\Lambda \gamma_1 n}{n_{\perp}}
\left[ 1 - \frac{\Lambda}{2}
\ln \left( \frac{|n|}{n_{\perp}} \right) \right]^{-1} \, , \label{delta2}
\end{eqnarray}
At high density $|n| \sim n_{\perp}$, the logarithmic term is
negligible and $U (n) \approx \Lambda \gamma_1 n/n_{\perp}$
is approximately linear in density. Note that the band gap
$U_g = |U|\gamma_1/\sqrt{U^2 + \gamma_1^2}$
tends to saturate $U_g \rightarrow \gamma_1$,
even if $|U| \gg \gamma_1$.
At low density, $|n| \ll n_{\perp}$, the logarithmic term
describing screening dominates and
$U_g \approx |U| \approx 2 \gamma_1 (|n|/n_{\perp}) / \ln ( n_{\perp}/|n| )$,
independent of the screening parameter $\Lambda$.

The expressions~(\ref{delta},\ref{delta2})
for $U ( n )$ take into account
screening due to low-energy electrons in $p_z$ orbitals
using a simplified Hamiltonian~(\ref{hsim}) while neglecting
other orbitals and the effects of
disorder \cite{zhang08,fogler10,nil07f,castro-peres07,mkh08,abergel11,min-abergel11,abergel11c,rutter11},
crystalline inhomogeneity \cite{min07} and electron-electron
exchange and correlation. Nevertheless,
there is generally good qualitative agreement of the dependence
of $U ( n )$ on density $n$ predicted by
equations~(\ref{delta},\ref{delta2})
with density functional theory calculations \cite{min07,gava}
and experiments
including photoemission \cite{ohta06,castro}
and infrared spectroscopy \cite{hen08,zhang08,kuz09a,mak09,zhang09,kuz09b}.
Note that the Hartree screening model has been generalised to describe
graphene trilayers and multilayers \cite{kosh09a,avet,avet2,kosh10a}.

\section{Transport properties}\label{s:tp}

\subsection{Introduction}

Bilayer graphene exhibits peculiar transport properties
due to its unusual band structure, described in section~\ref{s:ebs},
where the conduction and valence bands
touch with quadratic dispersion.
Transport characteristics
and the nature of conductivity near the Dirac point
were probed experimentally
\cite{novo05,zhang05,novo06,gor07,castro,oostinga,Moro08,Feld09a,xiao10}
and investigated theoretically
\cite{Kosh06,Nils06,Katz06,kats07,Cser07a,Snym07,Cser07b,Kech07,adam08,dassarma10,trush10,hwang10b,ferreira11}.
Neglecting trigonal warping, the minimal conductivity
is predicted to be $8e^2/(\pi h)$ \cite{Kosh06,Cser07a,Snym07}, twice the value in monolayer
graphene, while, in the presence of trigonal warping,
it is larger, $24e^2/(\pi h)$ \cite{Kosh06,Cser07b},
because of multiple Fermi surface pockets at low energy, section~\ref{s:tw}.

For a detailed review of the electronic transport properties of graphene
monolayer and bilayers, see Ref.~\cite{peres10,dassarma11}.
The characteristics of bilayer graphene in the presence of short-ranged defects
and long-ranged charged-impurities have been
calculated \cite{Kosh06,Nils06,kats07,nil08,adam08,dassarma10,trush10,hwang10b,ferreira11}
and it is predicted that the conductivity has an approximately linear dependence
on density at typical experimental densities \cite{dassarma10}.
At interfaces and potential barriers, conservation of the pseudospin
degree of freedom may influence electronic transmission \cite{kat06,poole10},
as in monolayer graphene \cite{che06,kat06},
including transmission at monolayer-bilayer interfaces \cite{nil-cneto07,nak10,koshino10,gonz11},
through multiple electrostatic barriers \cite{barbier09}, or magnetic barriers \cite{ram09,ghosh12}.
Inducing interlayer asymmetry and a band gap using an external gate \cite{mcc06a,mcc06b},
described in section~\ref{s:tbg}, may be used to tune transport properties \cite{nil-cneto07,fiori09}.
Interlayer asymmetry may also be viewed
as creating an out-of-plane component of pseudospin
and interfaces between regions of opposite polarity have
attracted theoretical attention due to the existence of one-dimensional
valley-polarised modes along the interface \cite{martin08,killi10,qiao11,zarenia11},
a pseudospintronic analogy of spin-valve devices for transport
perpendicular to the interface \cite{sanjose09,li10b},
electronic confinement \cite{xavier10},
and valley-dependent transmission \cite{schom10}.

In the following we describe two different models of the conductivity
of bilayer graphene at low energy. The first is for ballistic transport in
a clean device of finite length that is connected to semi-infinite leads,
described using wave matching to calculate the transmission probability and,
then, the conductance.
The second model describes the conductivity of a disordered, infinite system
using the Kubo formula and the self-consistent Born approximation
to describe scattering from the disordered potential.
Although the two models are quite different, both predict the minimal conductivity
to be $8e^2/(\pi h)$ \cite{Kosh06,Cser07a,Snym07}.
Finally, in section \ref{ss:localisation}, we describe localisation effects.

\subsection{Ballistic transport in a finite system}

Ballistic transport in a finite, mesoscopic bilayer graphene nanostructure has
been modelled in a number of papers \cite{Katz06,kat06,Cser07a,Snym07,Cser07b,sanjose09,li10b,schom10}.
Here, we follow a wave-matching approach of Snyman and Beenakker \cite{Snym07}.
For bilayer, as compared to monolayer, there is a new length scale
$\ell_1 = \hbar v /\gamma_1$ characteristic of the
interlayer coupling. Here, $v = \sqrt{3} a \gamma_0 / 2\hbar$ is the band velocity
of monolayer graphene, so $\ell_1 = (\sqrt{3}/2)(\gamma_0 / \gamma_1)a
\approx 18\,$\AA$\,$ is several times longer \cite{Snym07} than the lattice constant
$a = 2.46\,$\AA$\,$\cite{saito}. For most situations, the sample size $L \gg \ell_1$,
and the device generally behaves as a (coupled) bilayer rather than two separate
monolayers \cite{Snym07}.

\begin{figure}[t]
\centerline{\epsfxsize=0.9\hsize \epsffile{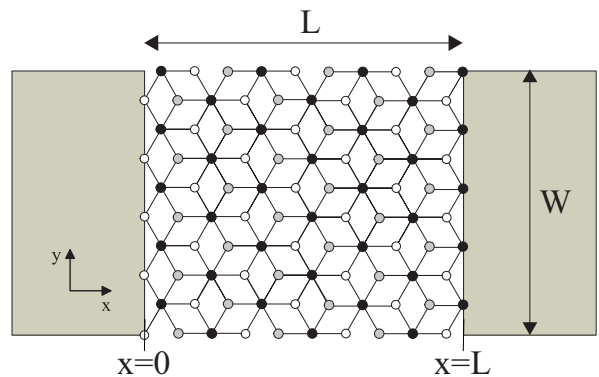}}
\caption{Two-probe bilayer graphene device with armchair edges,
width $W$ and length $L$. The rectangular shaded regions represent
the ends of semi-infinite leads.}
\label{fig:ballisticsetup}
\end{figure}

We consider a two-probe geometry with armchair edges as shown in Fig.~\ref{fig:ballisticsetup}.
There is a central mesoscopic bilayer region, width $W$ and length $L$, connected to a
left and right lead. This orientation is rotated by $90^{\circ}$ as compared to that
described in section~\ref{s:ebs} so the corners of the Brillouin zone
are located at wavevectors ${\mathbf K}_{\xi} = \xi ( 0 , 4 \pi/3a )$.
In terms of wavevector measured from the centre of the valley,
{\em i.e.}, $k_y \rightarrow k_y + \xi 4 \pi / 3a$, then the Hamiltonian~(\ref{Hbmin})
in basis $A1,B1,A2,B2$ may be written as
\begin{eqnarray}
H_{\mathrm{b}} &=& \left(
      \begin{array}{cccc}
        U & v \pi^{\dagger} & 0 & 0 \\
        v \pi & U & \gamma_1 & 0  \\
        0 & \gamma_1 & U & v \pi^{\dagger} \\
        0 & 0 & v \pi & U \\
      \end{array}
    \right)  \label{harmchair} ,
\end{eqnarray}
where $\pi = -i \hbar (k_x + i\xi k_y)$, $\pi^{\dagger} = i \hbar (k_x - i\xi k_y)$,
and $U$ is the on-site energy which describes the doping of the bilayer.
For simplicity, we include only the main interlayer coupling term $\gamma_1$ of the
orbitals on the dimer $B1$ and $A2$ sites.
It is assumed that the transverse wavevector $k_y$ is real and it is conserved at the
interfaces between the bilayer and the leads. The Hamiltonian~(\ref{harmchair})
shows there are two values of the longitudinal wavevector $k_x$ for given $U$, $k_y$
and energy $\varepsilon$, which we denote as $k_{+}$ and $k_{-}$:
\begin{eqnarray}
\hbar v k_{\pm} = \sqrt{ (\varepsilon-U)^2 \pm \gamma_1 (\varepsilon-U) - \hbar^2 v^2 k_y^2 } \, . \label{kpm}
\end{eqnarray}
Left- $\Phi_{\pm}^{L}$ and right- $\Phi_{\pm}^{R}$ moving wave functions may be written as
\begin{eqnarray}
\Phi_{\pm}^{L} &=& N_{\pm} \left(
                           \begin{array}{c}
                             \mp i \hbar v (-k_{\pm}-i\xi k_y) \\
\mp (\varepsilon-U) \\
                             (\varepsilon-U) \\
- i \hbar v (-k_{\pm}+i\xi k_y)  \\
                           \end{array}
                         \right)
e^{-i k_{\pm} x + i k_y y}  \, , \\
\Phi_{\pm}^{R} &=& N_{\pm} \left(
                           \begin{array}{c}
                             \mp i \hbar v (k_{\pm}-i\xi k_y) \\
\mp (\varepsilon-U) \\
                             (\varepsilon-U) \\
- i \hbar v (k_{\pm}+i\xi k_y)  \\
                           \end{array}
                         \right)
e^{i k_{\pm} x + i k_y y} \, ,
\end{eqnarray}
where normalisation $N_{\pm} = [4 W (\varepsilon-U) k_{\pm}]^{-1/2}$ for unit current.

The aim is to describe a mesoscopic bilayer region of finite length $L$ connected to macroscopic leads.
In order to mimick macroscopic, metallic contacts, there should be many propagating modes
in the leads that overlap with the modes in the central bilayer region.
If this is the case, then the value of minimal conductance should not depend on the
particular model used for the leads, {\em e.g.}, square lattice or graphene lattice,
as demonstrated for monolayer graphene \cite{schomerus07}.
Note, this will yield quite different results from a model with a bilayer lead
at the same level of doping as the central region; then, the system is effectively an infinite
system, not a finite, mesoscopic conductor.

Snyman and Beenakker \cite{Snym07} modelled the leads as heavily-doped bilayer graphene,
generalising an approach developed for monolayer graphene \cite{two06}.
In this way, there are many conducting modes present in the leads
and it is possible to simply use matching of the bilayer wave functions
at the interface between the central region and the leads.
In particular, the leads are modelled as bilayer graphene with on-site energy $U = - U_{\infty}$
where $U_{\infty} > 0$ and $U_{\infty} \gg \left\{ |\varepsilon| , \gamma_1 \right\}$. Then,
in the leads, $k_{+} \approx k_{-} \approx U_{\infty} / (\hbar v)$ and wave functions
$\psi_{\mathrm{left},\pm}$, $\psi_{\mathrm{right},\pm}$, in the left ($x<0$) and right ($x>L$) leads
may be written as
\begin{eqnarray}
\psi_{\mathrm{left},\pm} &=& \left[
\left(
                           \begin{array}{c}
                             \mp i \\
\mp 1 \\
                             1 \\
- i \\
                           \end{array}
                         \right)
e^{i U_{\infty} x / \hbar v}
+ r_{+}^{\pm} \left(
                           \begin{array}{c}
                             i \\
-1 \\
                             1 \\
i \\
                           \end{array}
                         \right)
e^{-i U_{\infty} x / \hbar v} \right. \nonumber \\
&& \left. \qquad \qquad
+ \, r_{-}^{\pm} \left(
                           \begin{array}{c}
                            -i \\
1 \\
                             1 \\
i \\
                           \end{array}
                         \right)
e^{-i U_{\infty} x / \hbar v}
\right]
e^{i k_y y} \, , \label{left-lead} \\
\psi_{\mathrm{right},\pm} &=& \left[
t_{+}^{\pm} \left(
                           \begin{array}{c}
                             -i \\
-1 \\
                             1 \\
-i \\
                           \end{array}
                         \right)
+ t_{-}^{\pm} \left(
                           \begin{array}{c}
                             i \\
1 \\
                             1 \\
-i \\
                           \end{array}
                         \right)
\right]
e^{i U_{\infty} (x-L) / \hbar v + i k_y y} \, . \nonumber
\end{eqnarray}
Here, a right-moving wave with unit flux corresponding to a state
with wavevector $k_{\pm} \approx U_{\infty} / (\hbar v)$ is injected from the left
lead [the first term in equation~(\ref{left-lead})]. Subsequently, there are two
left-moving waves that have been reflected with amplitudes
$r_{+}^{\pm}$ and $r_{-}^{\pm}$, and two right-moving waves
are transmitted to the right, with amplitudes $t_{+}^{\pm}$ and $t_{-}^{\pm}$.

\begin{figure}[t]
\centerline{\epsfxsize=0.8\hsize \epsffile{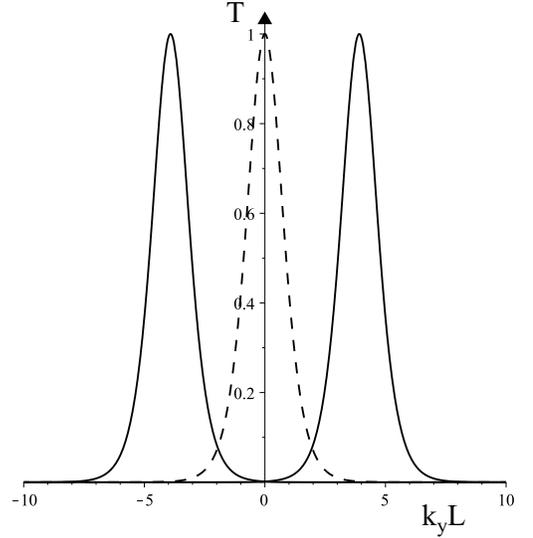}}
\caption{The transmission coefficients $T_{\pm}$ of bilayer graphene~(\ref{transcoeffs})
for $L = 50 \ell_1$ \cite{Snym07} (solid lines). The transmission coefficient of monolayer
graphene \cite{two06} is shown in the centre (dashed line).}
\label{fig:transmission}
\end{figure}

At the charge-neutrality point $\varepsilon=U=0$ in the central bilayer region,
the wave functions are evanescent with imaginary longitudinal wavevector,
equation~(\ref{kpm}). Two states with $k_x = -i \xi k_y$ have finite amplitude
only on the $A1$, $A2$ sites and two with $k_x = i \xi k_y$ have finite amplitude
only on the $B1$, $B2$ sites:
\begin{eqnarray*}
\psi_{\mathrm{centre},\pm} &=& \left[
c_{1}^{\pm} \left(
                           \begin{array}{c}
                             1 \\
                             0 \\
                             0 \\
                             0 \\
                           \end{array}
                         \right) e^{\xi k_y y}
+ c_{2}^{\pm} \left(
                           \begin{array}{c}
                             \gamma_1 x / \hbar v \\
                             0 \\
                             1 \\
                             0 \\
                           \end{array}
                         \right) e^{\xi k_y y} \right. \\
&& \!\!\! \!\!\! \!\!\! \!\!\! \!\!\! \!\!\! \left.
+ c_{3}^{\pm} \left(
                           \begin{array}{c}
                             0 \\
 1 \\
                             0 \\
-\gamma_1 x / \hbar v \\
                           \end{array}
                         \right) e^{-\xi k_y y}
+ c_{4}^{\pm} \left(
                           \begin{array}{c}
                             0 \\
0 \\
                             0 \\
1 \\
                           \end{array}
                         \right) e^{-\xi k_y y}
\right]
e^{i k_y y} \, .
\end{eqnarray*}
For each incoming mode $\pm$ from the left lead [the first term in equation~(\ref{left-lead})],
there are eight unknown amplitudes $c_{1}^{\pm}$, $c_{2}^{\pm}$, $c_{3}^{\pm}$, $c_{4}^{\pm}$,
$r_{+}^{\pm}$, $r_{-}^{\pm}$, $t_{+}^{\pm}$, $t_{-}^{\pm}$. Continuity of the
wave functions at the interfaces $x=0$ and $x=L$ between the central region and the leads
provides eight simultaneous equations:
\begin{eqnarray}
\mp i + i r_{+}^{\pm} - i r_{-}^{\pm} &=& c_{1}^{\pm} \, , \\
\mp 1 - r_{+}^{\pm} + r_{-}^{\pm} &=& c_{3}^{\pm} \, , \\
1 + r_{+}^{\pm} + r_{-}^{\pm} &=& c_{2}^{\pm} \, , \\
- i + i r_{+}^{\pm} + i r_{-}^{\pm} &=& c_{4}^{\pm} \, , \\
- i t_{+}^{\pm} + i t_{-}^{\pm} &=& c_{1}^{\pm} e^{\xi k_y L} + c_{2}^{\pm} \frac{L}{\ell_1} e^{\xi k_y L} \, , \\
- t_{+}^{\pm} + t_{-}^{\pm} &=& c_{3}^{\pm} e^{-\xi k_y L} \, ,\\
t_{+}^{\pm} + t_{-}^{\pm} &=& c_{2}^{\pm} e^{\xi k_y L} \, , \\
- i t_{+}^{\pm} - i t_{-}^{\pm} &=& -c_{3}^{\pm} \frac{L}{\ell_1} e^{-\xi k_y L} + c_{4}^{\pm} e^{-\xi k_y L} \, ,
\end{eqnarray}
where $\ell_1 = \hbar v /\gamma_1$.
Solving yields the transmission matrix
\begin{eqnarray*}
{\bf t} &=& \left(
            \begin{array}{cc}
              t_{+}^{+} & t_{+}^{-} \\
              t_{-}^{+} & t_{-}^{-} \\
            \end{array}
          \right) \\
&=& \frac{2i}{2 + (L/\ell_1)^2 + 2 \cosh \left( 2 k_y L \right)} \\
&& \!\!\!\!\!\! \times
\left(
            \begin{array}{cc}
              \left( L/\ell_1 - 2 i \right) \cosh \left( k_y L \right) & -\left( L/\ell_1 \right) \sinh \left( \xi k_y L \right) \\
              \left( L/\ell_1 \right) \sinh \left( \xi k_y L \right) & -\left( L/\ell_1 + 2 i \right) \cosh \left( k_y L \right) \\
            \end{array}
          \right) \, .
\end{eqnarray*}
The transmission probability \cite{Snym07} is determined by the eigenvalues $T_{\pm}$ of the product ${\bf t}{\bf t}^{\dagger}$:
\begin{eqnarray}
T_{\pm} &=& \frac{1}{\cosh^2 \left( \xi k_y L \mp k_c L \right)} \, , \label{transcoeffs} \\
k_c L &=& \ln \left[ \frac{L}{2\ell_1} + \sqrt{1 + \left( \frac{L}{2\ell_1} \right)^2} \right] \, ,
\end{eqnarray}
The transmission coefficients $T_{\pm}$ have the same form as the
transmission in monolayer graphene $T = 1 / \cosh^2 (k_y L)$  \cite{two06} but shifted by
the parameter $k_c$, as shown in figure~\ref{fig:transmission}.

The conductance $G$ may be determined using the Landauer-B\"uttiker formula \cite{landauer,buttiker}
\begin{eqnarray}
G = \frac{g_v g_s e^2}{h} \mathrm{Tr} \left( {\bf t}{\bf t}^{\dagger} \right) ,
\end{eqnarray}
where the factor of $g_v g_s$ accounts for valley and spin degeneracy.
For a short, wide sample whose width $W$ exceeds its length $L$, $W \gg L$,
the transverse wavevector may be assumed to be continuous and
\begin{eqnarray}
G = \frac{g_v g_s e^2}{h} \frac{W}{2\pi} \int_{-\infty}^{\infty} \left( T_{+} + T_{-} \right) d k_y
= \frac{2 g_v g_s e^2}{h} \frac{W}{\pi L} \, .
\end{eqnarray}
Thus, the minimal conductivity
$\sigma = {G L}/{W} = {8e^2}/({\pi h})$ is twice as large as in the monolayer.
In a similar way, it is possible to determine the Fano factor of shot noise which takes the same
value $1/3$ \cite{Snym07} as in monolayer graphene \cite{two06}.
Transmission via evanescent modes in graphene has been described as {\em pseudodiffusive}
because the Fano factor takes the same value as in a diffusive metal \cite{two06}.

\subsection{Transport in disordered bilayer graphene}

\subsubsection{Conductivity}

When the Fermi energy $\vare_F$ is much larger than
the level broadening caused by the disorder potential,
the system is not largely different from a conventional metal,
and the conductivity is well described by
Boltzmann transport theory.
However, this approximation inevitably breaks down
at the Dirac point,
where even the issue of whether the system is metallic or insulating is nontrivial.
To model electronic transport at the charge neutrality point,
we need a refined approximation that properly includes the finite level broadening.
Here, we present a conductivity calculation
using the self-consistent Born approximation (SCBA) \cite{Kosh06}.
We define the Green's function as $G(\vare) = (\vare - {\cal H})^{-1}$.
The Green's function averaged over the impurity configurations
satisfies the Dyson's equation
\begin{eqnarray}
 \langle G_{\alpha, \alpha'}(\vare) \rangle &=&
\delta_{\alpha, \alpha'}G^{(0)}_{\alpha}(\vare) \nonumber \\
&& + \, G^{(0)}_{\alpha}(\vare)\sum_{\alpha_1}
\Sigma_{\alpha,\alpha_1}(\vare) \langle G_{\alpha_1, \alpha'}(\vare) \rangle ,  \label{eq_dyson}
\end{eqnarray}
where $\langle\,\rangle$ is an average over configurations of the disorder potential,
$\alpha$ is an eigenstate of the ideal Hamiltonian ${\cal H}_0$,
and $G_{\alpha}^{(0)} = (\vare - \vare_\alpha)^{-1}$,
with $\vare_\alpha$ being the eigenenergy of the state $\alpha$
in ${\cal H}_0$.
In SCBA, the self-energy  is given by \cite{Shon_and_Ando_1998a}
\begin{equation}
\Sigma_{\alpha,\alpha'} (\vare) =
\sum_{\alpha_1, \alpha_1'}
\langle
U_{\alpha,\alpha_1}U_{\alpha'_1,\alpha'}
\rangle
\langle
G_{\alpha_1,\alpha'_1}(\vare)
\rangle.
\label{eq_sigma}
\end{equation}
The equations (\ref{eq_dyson}) and (\ref{eq_sigma}) need to be solved self-consistently.
The conductivity is calculated using the Kubo formula,
\begin{equation}
 \sigma(\vare) = g_vg_s\frac{\hbar e^2}{2\pi\Omega} {\rm Re} {\rm Tr}
\left[v_x \av{G^R} \tilde{v}_x^{RA} \av{G^A}
-v_x \av{G^R} \tilde{v}_x^{RR} \av{G^R}
\right],
\end{equation}
where $G^R=G(\vare+i0)$ and $G^A=G(\vare-i0)$ are retarded and advanced Green's functions,
$v_x = \partial {\cal H}_0 / \partial p_x$ is the velocity operator,
and $g_vg_s$ accounts for summation over valleys and spins.
$\tilde{v}_x^{RA}$ and $\tilde{v}_x^{RR}$
the velocity operators containing the vertex correction,
defined by $\tilde{v}_x^{RA} =  \tilde{v}_x(\vare+i0,\vare-i0)$
and $\tilde{v}_x^{RR} =  \tilde{v}_x(\vare+i0,\vare+i0)$ with
\begin{equation}
 \tilde{v}_x(\vare,\vare') =  v_x + \av{UG(\vare)\tilde{v}_xG(\vare')U}.
\label{eq_vertex}
\end{equation}
In SCBA, $\tilde v_x$ should be calculated in the ladder approximation.

For the disorder potential, we assume a short-ranged potential within each valley,
\begin{equation}
 U =
\sum_i u_i \delta(\mathbf{r} - \mathbf{r}_i)
\left(
\begin{array}{cccc}
1 & 0 & 0 & 0\\
0 & 1 & 0 & 0\\
0 & 0 & 1 & 0\\
0 & 0 & 0 & 1
\end{array}
\right).
\label{eq_V}
\end{equation}
and neglect intervalley scattering between $K_{+}$ and $K_{-}$.
This situation is realized when the
length scale of the scattering potential is larger than atomic scale
but much shorter than the Fermi wave length.
We assume an equal amount of positive and negative
scatterers $u_i = \pm u$ and a total density of scatterers per unit area $n_{\rm imp}$.

The SCBA formulation is applied to
the low-energy Hamiltonian equation~(\ref{heff1}),
with the trigonal warping effect due to $\gamma_3$ included
\cite{Kosh06}. Energy broadening due to the disorder potential
is characterised by
\begin{equation}
 \Gamma = \frac{\pi}{2} n_{\rm imp} u^2 \frac{m^*}{2\pi\hbar}.
\label{eq_gamma}
\end{equation}
When $\Gamma \gsim \vare_{L}$,
{\em i.e.}, the disorder is strong enough to
smear out the fine band structure of the trigonal warping,
we can approximately solve the SCBA equation in an analytic form.
The self-energy, equation~(\ref{eq_sigma}),
becomes diagonal with respect to index $\alpha$
and it is a constant,
\begin{equation}
 \Sigma(\vare+i0) \approx -i \Gamma.
\end{equation}
Then, the conductivity at the Fermi energy $\vare$ is written as
\begin{equation}
  \sigma(\vare) \approx g_vg_s\frac{e^2}{\pi^2\hbar}
\frac{1}{2}\left[
1+
\left(\frac{|\vare|}{\Gamma}
+ \frac{\Gamma}{|\vare|} \right)
\arctan \frac{|\vare|}{\Gamma}
+ \frac{4\pi\vare_{L}}{\Gamma}
\right].
\label{eq_cond_scba}
\end{equation}
The third term in the square bracket arises from the vertex correction
due to the trigonal warping effect, and $\vare_L$ is given by equation~(\ref{eq_lifshitz}).
For high energies $|\vare|\gg \Gamma$, $\sigma$ approximates as
\begin{equation}
 \sigma(\vare) \approx g_vg_s\frac{e^2}{\pi^2\hbar}
\frac{\pi}{4} \frac{|\vare|}{\Gamma},
\label{eq_cond_scba_high}
\end{equation}
which increases linearly with energy. The value at zero energy becomes
\begin{equation}
 \sigma(0) = g_vg_s\frac{e^2}{\pi^2\hbar}
\left(
1+ \frac{2\pi\vare_{L}}{\Gamma}
\right).
\label{eq_cond_scba_0}
\end{equation}
In the strong disorder regime $\Gamma \gg 2\pi\vare_{L}$, the correction
arising from trigonal warping vanishes and the conductivity approaches
the universal value $g_vg_se^2/(\pi^2\hbar)$ \cite{Kosh06,Cser07a}
which is twice as large as that in monolayer graphene in the same approximation.
In transport measurements of suspended bilayer graphene \cite{Feld09a},
the minimum conductivity was estimated to be about $10^{-4}$S, which
is close to $g_vg_se^2/(\pi^2\hbar)$.

\begin{figure}
\begin{center}
\centerline{\epsfxsize=0.7\hsize \epsffile{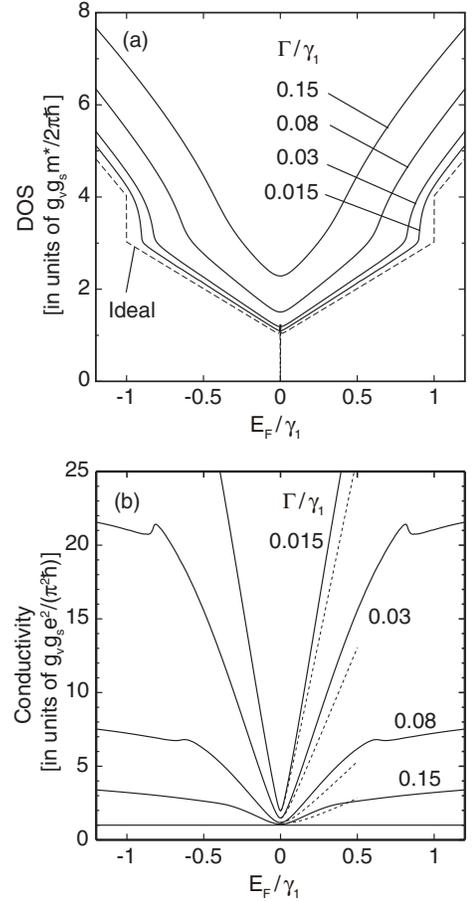}}
\caption{
(a) Calculated density of states and (b) conductivity
as a function of energy for different disorder strength $\Gamma$ \cite{koshinonjp09}.
In (b), broken curves show the results of the low-energy two-band model, and
the horizontal line indicates the universal conductivity $g_vg_s e^2/(\pi^2\hbar)$.}
\label{fig_bi_cond}
\end{center}
\end{figure}

The 2$\times$2 (two-band) model works well at low energy,
but it is not expected to be valid in the strong disorder regime when
mixing to higher energy bands is considerable.
To see this, we numerically solved
SCBA equation for the original 4$\times$4 (four-band) Hamiltonian.
Figure~\ref{fig_bi_cond}(a) and (b) show the density of states (DOS)
and conductivity, respectively, for several disorder strengths \cite{koshinonjp09}.
In (b) the results for the 2$\times$2 model in equation~(\ref{eq_cond_scba})
are expressed as broken curves.
In (a), we observe that the DOS at zero energy
is significantly enhanced because states at high energies
are shifted toward the Dirac point by the disorder potential.
However, the zero-energy conductivity barely
shifts from that of the 2$\times$2 model [equation~(\ref{eq_cond_scba_0})],
even for strong disorder such as $\Gamma/\gamma_1 = 0.15$,
where the DOS at zero energy becomes nearly twice as large as in the 2$\times$2 model.
For higher energy $|\vare| > \Gamma$, the conductivity increases linearly with
$|\vare|$ in qualitative agreement with equation~(\ref{eq_cond_scba}) of the 2$\times$2 model,
while the gradient is generally steeper.
The conductivity has a small dip at the higher band edge around $|\vare| \sim \gamma_1$,
because the frequency of electron scattering is strongly enhanced by the higher band states.

The SCBA calculation was recently extended for long-range scatterers \cite{Ando_2011a}.
It was shown that the conductivity at zero energy is not universal but depends on
the degree of disorder for scatterers with long-range potential,
similar to monolayer graphene \cite{Noro_et_al_2010a}.

\subsubsection{Localisation effects}\label{ss:localisation}

The SCBA does not take account of some quantum corrections, such as those included
explicitly in weak localisation.
In graphene, the presence of spin-like degrees of freedom related to sublattices and
valleys, as well as real electronic spin itself, creates the possibility of a
rich variety of quantum interference behaviour. Weak localisation \cite{WL,LarkinWAL}
is a particularly useful probe because it is sensitive to elastic scattering that causes
relaxation of the sublattice pseudospin and valley `spin'. In the absence of symmetry-breaking
scattering processes, conservation of pseudospin in monolayer graphene tends to suppress
backscattering \cite{ando98,mceuen99}, and the interference of chiral electrons
would be expected to result in antilocalization \cite{AndoWL}. However, intravalley symmetry-breaking relaxes the pseudospin
and suppresses anti-localisation, while intervalley scattering
tends to restore conventional weak localisation \cite{WLman,guinea06wl,WLmono,WLreview,falko07},
as observed experimentally \cite{WLman,heersche07,wu07,tik08,ki08,tik09,lara11}.
Nevertheless, anti-localisation has been observed at high temperature \cite{tik09,mcc09}
when the relative influence of symmetry-breaking disorder is diminished, and its presence
has also been predicted at very low temperature \cite{imura09,imura10,mcc12b} when
spin-orbit coupling may influence the spin of the interfering electrons.

In bilayer graphene, the pseudospin turns twice as quickly in the graphene
plane as in a monolayer, no suppression of backscattering is expected
and the quantum correction should be conventional weak localisation \cite{WLbilayer,WLreview,falko07}.
However, the relatively strong trigonal warping of the Fermi line
around each valley (described in section~\ref{s:tw})
can suppress localisation unless intervalley scattering
is sufficiently strong \cite{WLbilayer}.
Experimental observations confirmed this picture \cite{gor07}, and it was possible to
determine the temperature and density dependence of relevant relaxation lengths
by comparing to the predicted magnetoresistance formula \cite{WLbilayer}.

Localisation has also been studied for gapped bilayer graphene
in the presence of interlayer potential asymmetry $U$ \cite{kosh08_deloc}.
It was shown that, as long as the disorder potential is long range and does
not mix $K_\pm$ valleys, gap opening inevitably causes
electron delocalisation somewhere between $U=0$ and $U=\infty$,
in accordance with the transition of quantum valley Hall conductivity,
{\em i.e.}, the opposite Hall conductivities associated with two valleys.
This is an analog of quantum Hall physics
but can be controlled purely by an external electric field
without any use of magnetic fields.

\section{Optical properties}
\label{sec_bi_optical}

\begin{figure}
\begin{center}
\centerline{\epsfxsize=0.7\hsize \epsffile{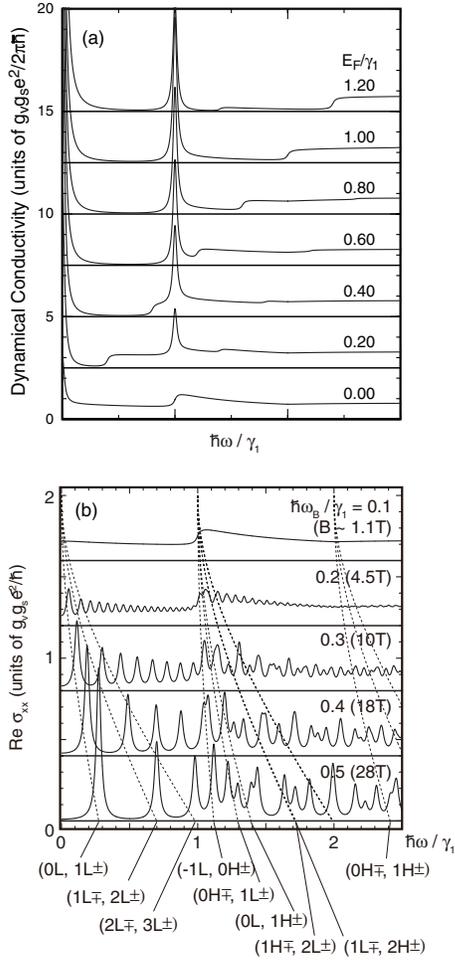}}
\caption{
Interband part of the dynamical conductivity of
bilayer graphene plotted against the frequency $\omega$,
in (a) zero magnetic field with different $\vare_F$'s \cite{Ando_and_Koshino_2009b}
(b) several magnetic fields with $\vare_F=0$ \cite{koshinoandoprb08}.
Dashed curves in (b) indicate the transition energies
between several Landau levels.}
\label{fig_bi_opt}
\end{center}
\end{figure}

The electronic structure of bilayer graphene
was probed by spectroscopic
measurements in zero magnetic field
\cite{ohta06,zhang08,li09,kuz09a,mak09,zhang09,kuz09b},
and also in high magnetic fields \cite{hen08}.
The optical absorption for perpendicularly incident light
is described by the dynamical conductivity
in a electric field parallel to the layers,
in both symmetric bilayers \cite{Nils06,Abergel_and_Falko_2007a,nicol08,Koshino_and_Ando_2009b}
and in the presence of
an interlayer-asymmetry gap \cite{nicol08}
For symmetric bilayer graphene,
this is explicitly estimated as
\cite{Nils06,Abergel_and_Falko_2007a,nicol08,Koshino_and_Ando_2009b}
\begin{eqnarray}
 {\rm Re} \, \sigma_{xx} (\omega)  &=&
\frac{g_vg_s}{16}\frac{e^2}{\hbar}
\left\{
\frac{\hbar\omega+2\gamma_1}{\hbar\omega+\gamma_1}
\theta(\hbar\omega - 2|\vare_F|)
\right.
\nonumber\\
&&
+\left(\frac{\gamma_1}{\hbar\omega}\right)^2
\left[
\theta(\hbar\omega -\gamma_1) +
\theta(\hbar\omega -\gamma_1- 2|\vare_F|)
\right]
\nonumber\\
&&
+
\frac{\hbar\omega-2\gamma_1}{\hbar\omega-\gamma_1}
\theta(\hbar\omega -2\gamma_1)
\nonumber\\
&&
\left.
+
\gamma_1
\log\left[\frac{2|\vare_F|+\gamma_1}{\gamma_1}\right]
\delta(\hbar\omega-\gamma_1)
\right\},
\label{eq_bi_dyn}
\end{eqnarray}
where $\vare_F$ is the Fermi energy and we assumed $|\vare_F| < \gamma_1$.
We label the four bands in order of descending energy as $1,2,3,4$.
The first term in equation~(\ref{eq_bi_dyn}) represents
absorption from band 2 to 3,
the second from 2 to 4 or from 1 to 3,
the third from 1 to 4,
and the fourth from 3 to 4 or from 1 to 2.
Figure \ref{fig_bi_opt} (a) shows some examples of calculated dynamical
conductivity ${\rm Re}\,\sigma_{xx}(\omega)$
with several values of the Fermi energy
\cite{Ando_and_Koshino_2009b}.
The curve for $\vare_F=0$ has essentially
no prominent structure except for a step-like
increase corresponding to transitions from 2 to 4.
With an increase in $\vare_F$, a delta-function peak
appears at $\hbar\omega = \gamma_1$, corresponding to
allowed transitions 3 to 4.

In a magnetic field, an optical excitation
by perpendicular incident light is only allowed
between the Landau levels with $n$
and $n\pm 1$ for arbitrary combinations of $\mu=H, \, L$
and $s=\pm 1$, since the matrix element of the
velocity operator $v_x$ vanishes otherwise.
Figure~\ref{fig_bi_opt} (b) shows some plots of
${\rm Re}\,\sigma_{xx}(\omega)$
in magnetic fields at $\vare_F = 0$ and zero temperature \cite{koshinoandoprb08}.
Dotted lines penetrating panels
represent the transition energies between
several specific Landau levels as a continuous function of $\hbar\omega_B$.
Every peak position behaves as a linear function of
$B \propto \hbar\omega_B^2$ in weak field
but it switches over to $\sqrt{B}$-dependence as the corresponding energy
moves out of the parabolic band region.
In small magnetic fields, the peak structure is
smeared out into the zero-field curve
more easily in the bilayer than in the monolayer,
because the Landau level spacing is narrower
in bilayer due to the finite band mass.

\section{Orbital magnetism}\label{s:om}

Graphite and related materials exhibit a strong orbital diamagnetism which
overcomes the Pauli spin paramagnetism.
Theoretically the diamagnetic susceptibility was calculated for graphite
\cite{mcclure56,McClure_1960a,Sharma_et_al_1974a},
graphite intercalation compounds \cite{Safran_and_DiSalvo_1979a,Safran_1984a,Blinowski_and_Rigaux_1984a,Saito_and_Kamimura_1986a},
as well as few-layer graphenes \cite{Koshino_and_Ando_2007a,Nakamura_and_Hirasawa_2008a}.
In particular, monolayer graphene has a strong diamagnetic singularity at Dirac point,
which is expressed as a Delta function in Fermi energy $\vare_F$
\cite{mcclure56,Koshino_and_Ando_2007b,Fukuyama_2007a,Nakamura_2007a,Ghosal_et_al_2007a}.
In the bilayer, the singularity is relaxed by the modification of the band structure
caused by the interlayer coupling as we will see in the following
\cite{Safran_1984a,Koshino_and_Ando_2007a}.

When the spectrum is composed of discrete Landau levels $\vare_n$,
the thermodynamical potential is generally written as
\begin{eqnarray}
\Omega = -\frac{1}{\beta} \frac{g_vg_s}{2\pi l_B^2}
\sum_{n}
\varphi(\varepsilon_n),
\label{eq_omega}
\end{eqnarray}
where $\beta = 1/k_B T$,
$\varphi(\varepsilon) = \log \big[ 1 + e^{-\beta(\varepsilon - \zeta)} \big]$
with $\zeta$ being the chemical potential.
In weak magnetic field, using the Euler-Maclaurin formula, the summation
in $n$ in equation~(\ref{eq_omega}) can be written as an integral in a
continuous variable $x$ with a residual term proportional to $B^2$.
The magnetization $M$ and the magnetic susceptibility $\chi$ can be calculated by
\begin{eqnarray}
\!\!\!\!\!\!\!\!\!\!\!\! M = -\Big(\frac{\partial \Omega}{\partial B}\Big)_\zeta, \,
\chi = \frac{\partial M}{\partial B} \Big|_{B=0}
= -\Big(\frac{\partial^2 \Omega}{\partial B^2}\Big)_\zeta \Big|_{B=0} .
\label{eq_chi_def}
\end{eqnarray}

For monolayer graphene, the susceptibility is
\cite{mcclure56,Koshino_and_Ando_2007b}
\begin{equation}
 \chi = -g_vg_s \frac{e^2 v^2}{6\pi c^2}
\int_{-\infty}^{\infty}
\left(-\frac{\partial f(\vare)}{\partial \vare}\right)
d\vare.
\end{equation}
At zero temperature, it becomes a delta function in Fermi energy,
\begin{equation}
\chi(\varepsilon_F) = -g_vg_s \frac{e^2 v^2}{6\pi c^2} \delta(\varepsilon_F).
\label{eq_chi_mono}
\end{equation}
The delta-function susceptibility of monolayer graphene
is strongly distorted by the interlayer coupling $\gamma_1$.
For the Hamiltonian of the symmetric bilayer graphene,
the orbital susceptibility is calculated as
\cite{Safran_1984a,Koshino_and_Ando_2007a}
\begin{eqnarray}
\chi(\varepsilon) = g_v g_s \frac{e^2 v^2}{4\pi c^2 \gamma_1}
 \theta(\gamma_1-|\varepsilon|)
\Big(
 \log \frac{|\varepsilon|}{\gamma_1} + \frac{1}{3}
\Big).
\label{eq_chi_bi}
\end{eqnarray}
The susceptibility diverges logarithmically at $\vare_F=0$, becomes
slightly paramagnetic near $|\vare_F| = \gamma_1$, and
vanishes for $|\vare_F| > \gamma_1$ where the higher
subband enters.
The integration of $\chi$ in equation~(\ref{eq_chi_bi}) over the Fermi
energy becomes $- g_v g_s e^2v^2 /(3\pi c^2)$
independent of $\gamma_1$, which is exactly twice as large as that of
the monolayer graphene, equation~(\ref{eq_chi_mono}).

\begin{figure}
\begin{center}
\centerline{\epsfxsize=0.7\hsize \epsffile{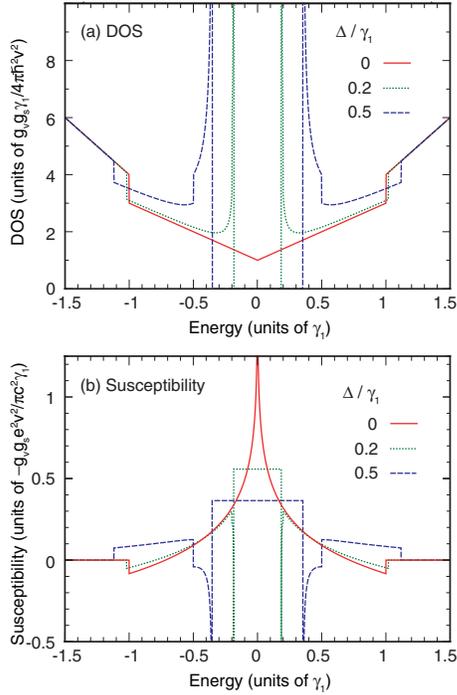}}
\caption{(a) Density of states, and (b) susceptibility of bilayer graphenes
with the asymmetry gap $U/\gamma_1 = 0$, 0.2, and 0.5 \cite{Koshino_and_Ando_2010a}.
In (b), the upward direction represents negative (i.e., diamagnetic)
susceptibility.
}
\label{fig_chi_bilayer}
\end{center}
\end{figure}

The susceptibility was also calculated in the presence
of interlayer asymmetry \cite{Koshino_and_Ando_2010a}.
Figure \ref{fig_chi_bilayer} (a) and (b) show
the density of states and
the susceptibility, respectively, for
bilayer graphene with $U=0$, 0.2, and 0.5.
The susceptibility diverges in the paramagnetic direction at the band
edges where the density of states also diverges.
This huge paramagnetism can be interpreted as the Pauli paramagnetism
induced by the valley pseudo-spin splitting
and diverging density of states \cite{Koshino_2011a}.
The susceptibility vanishes in the energy region
where the higher subband enters.

\section{Phonons and strain}\label{s:pas}

\subsection{The influence of strain on electrons in bilayer graphene}

Deformation of a graphene sheet couples to the electronic system
and modifies the low-energy Hamiltonian. In monolayer graphene,
static changes in distance and angles of the atomic bonds
can be described as effective scalar or vector potentials
in the Dirac Hamiltonian \cite{suzuuraando02,cnreview}.
Bilayer graphene has extra degrees of freedom in deformation
associated with the presence of two layers.
It was shown that a band gap can be opened by giving different distortions to the two layers
or by pulling the two layers apart in the perpendicular direction
\cite{nanda09,choi10,verberck12}.

Rather than produce a band gap, it has been predicted that homogeneous
lateral strain in bilayer graphene can produce a change in topology of the band structure
at low energy \cite{mucha11,son11,mariani11}. This deformation
causes tight-binding parameters $\gamma_0$ and $\gamma_3$ to become dependent
on the hopping direction and produces an additional term in the
low-energy two-band Hamiltonian~(\ref{heff1})
\begin{eqnarray}
{\hat{h}}_{s} &=& \left(
\begin{array}{cc}
0 & w \\
w^{\ast} & 0
\end{array}
\right) \, , \label{hstrain}
\end{eqnarray}
where parameter $w = | w | e^{i\xi\theta}$ depends on the microscopic details of
the deformation; it is non-zero only when the role of skew interlayer
coupling $\gamma_3$ is taken into account \cite{mucha11}.
It is estimated that $1\%$ strain would give $\left| w \right| \sim 6\,$meV \cite{mucha11}.
Taken with the quadratic term ${\hat{h}}_{0}$ in the Hamiltonian~(\ref{heff1}),
the low-energy bands with energy $E = \pm \varepsilon_1$ are given by
\begin{eqnarray}
\varepsilon_1^2 = \left| w \right|^2 - \frac{\left| w \right| p^2}{m}
\cos \left( 2 \varphi + \theta \right)
+ \left( \frac{p^2}{2m} \right)^2 \, .
\end{eqnarray}
This describes a Lifshitz transition at energy $\varepsilon_L = \left| w \right|$,
below which there are two Dirac points in the vicinity of each Brillouin zone corner  \cite{mucha11,son11,mariani11},
centred at momentum $p \approx \sqrt{2m\left| w \right|} = \sqrt{\left| w \right|\gamma_1}/v$
and angles $\varphi = - \theta /2$ and $\varphi = \pi - \theta /2$.
In general, there should be an interplay between terms ${\hat{h}}_{s}$, equation~(\ref{hstrain}),
and ${\hat{h}}_{w}$, equation~(\ref{heff1}), leading to the possibility of employing
strain to annihilate two Dirac points and, thus, change the low energy topology of the bands from
four to two Dirac points \cite{mucha11}.

The presence of two Dirac points would cause zero-energy Landau levels to be eightfold
degenerate; an experimental signature of this state is predicted to be
the persistence of filling factor $\nu = \pm 4$ in the low-field quantum Hall effect \cite{mucha11}.
This contrasts with the Lifshitz transition that would occur in the presence of
parameter $\gamma_3$ without strain when there are four Dirac points,
section~\ref{s:tw}, giving a degeneracy of sixteen and $\nu = \pm 8$ at low fields.
In both cases, Berry phase $2\pi$ is conserved:
two Dirac cones with Berry phase $\pi$ each \cite{mucha11}
or four Dirac cones with three of $\pi$ and one of $-\pi$ \cite{manes07,mik08}.
It has also been predicted that the presence of the Lifshitz transition
will be noticeable in the low-energy conductivity at zero magnetic field \cite{Kosh06,Cser07b},
and the particular case of two Dirac points in the presence of strain
has recently been analysed, too \cite{grad12}.
Note that the effect of lateral strain on the low-energy topology of
the band structure is qualitatively similar to that of
a gapless nematic phase which possibly arises as
the result of electron-electron interactions in bilayer
graphene \cite{vafek10a,lem10,throck12,lem12}.

\subsection{Phonons in bilayer graphene}

Raman spectroscopy has been a valuable tool in probing the behaviour of phonons in
graphite \cite{reich-thomsen04} and it may be used
to determine the number of layers in multilayer graphene \cite{ferrari06},
differentiating between monolayer and bilayer.
For an in-depth review of Raman spectroscopy of graphene including bilayer graphene see,
for example, Refs.~\cite{ferrari-review,mal-pim09}.
The phonon spectrum of monolayer graphene has been calculated using
a tight-binding force-constant model with parameters fit to Raman data \cite{grun02},
and with density functional theory \cite{dubay03,maultzsch04,wirtz04,mounet05}.
There are three acoustic (A) and three optical (O) branches consisting of longitudinal (L)
and transverse (T) in-plane modes as well as out-of-plane (Z) modes.
At the zone centre (the $\Gamma$ point), the TA and LA modes display linear dispersion
$\omega \sim q$ but the ZA mode is quadratic $\omega \sim q^2$.
The ZO mode is at $\sim 890\,$cm$^{-1}$ \cite{dubay03,wirtz04}, and the LO and TO modes
are degenerate (at $\sim 1600\,$cm$^{-1}$).
At the $K$ point, the ZA and ZO modes ($\sim 540\,$cm$^{-1}$) and the LA and LO modes
($\sim 1240\,$cm$^{-1}$) are degenerate,
with TA modes at $\sim 1000\,$cm$^{-1}$ and TO at $\sim 1300\,$cm$^{-1}$.
For undoped graphene, owing to strong electron-phonon coupling, the highest optical modes at the
$\Gamma$ and $K$ point ({\em i.e.} the LO mode at the
$\Gamma$ point and the TO mode at the $K$ point) display Kohn anomalies \cite{kohn59,pisc04,lazzeri06}
whereby the phonon dispersion $\omega (q)$ has an almost linear slope as observed, for example,
in inelastic x-ray measurements of graphite \cite{maultzsch04,lazzeri06}.
As graphene is a unique system in which the electron or hole concentration
can be tuned by an external gate voltage, it was realised \cite{ando06,castroneto07} that the change in electron
density would also influence the behaviour of the optical phonons through
electron-phonon coupling and, in particular, a logarithmic singularity in their
dispersion was predicted \cite{ando06} when the Fermi energy $\varepsilon_F$ is half of the energy of the optical phonon
$|\varepsilon_F| = \hbar \omega /2$. Subsequently, such tuning of phonon frequency
and bandwidth by adjusting the electronic density was observed in monolayer graphene through
Raman spectroscopy \cite{pisana07,yan-zhang07}.

The behaviour of phonons in bilayer graphene has been observed experimentally through Raman spectroscopy
\cite{mal07,yan08,malard08,das,yan09,bruna10,garcia10,tan-han12,lui12}
and infrared spectroscopy \cite{kuz09c,tang10},
with particular focus on optical phonon anomalies and the influence of gating.
Generally, the phonon spectrum of bilayer graphene, which has been calculated using density functional theory \cite{yan-ruan08,saha08}
and force-constant models \cite{jiang-tang08,michel-verberck08},
is similar to that of monolayer.
Near the $\Gamma$ point there are additional low-frequency modes.
There is a doubly-degenerate rigid shear mode at $\sim 30\,$cm$^{-1}$ \cite{jiang-tang08,michel-verberck08}
observed through Raman spectroscopy \cite{tan-han12} and
an optical mode at $\sim 90\,$cm$^{-1}$
which arises from relative motion of the layers in the vertical direction (perpendicular to the layer plane),
known as a layer-breathing mode \cite{kiti05} and observed through Raman spectroscopy \cite{lui12}.
At the $\Gamma$ point, interlayer coupling causes the LO/TO modes to split into two doubly-degenerate branches
where the higher (lower) frequency branch corresponds to symmetric `in-phase' (antisymmetric `out-of-phase')
relative motion of atoms on the two layers (in the in-plane direction).
Analogously to the monolayer, it was predicted that these optical phonons would
be affected by electron-phonon coupling, with a logarithmic singularity in the
dispersion of the symmetric modes when the Fermi energy $\varepsilon_F$
is equal to half of the optical phonon frequency \cite{ando07},
and hybridisation of the symmetric and antisymmetric modes in the
presence of interlayer potential asymmetry \cite{andokoshino09,gava09}.
Experimentally, this anomalous phonon dispersion has been observed through
Raman spectroscopy \cite{yan08,malard08,das,yan09,bruna10} including
the evolution of two distinct components in the Raman $G$ band for
non-zero interlayer asymmetry \cite{malard08,yan09,bruna10}.
The Raman spectrum has also been studied for bilayer graphene in the
presence of Landau levels in a magnetic field \cite{ando07,mucha10c}.

\subsection{Optical phonon anomaly}

In the following, we describe the anomalous optical phonon spectrum in bilayer graphene
taking into account electron-phonon coupling \cite{ando07,andokoshino09}.
Theoretically, it was shown that a continuum model works well in describing
long-wavelength acoustic phonons \cite{suzuuraando02} and optical phonons \cite{ishikawaando06}
in graphene, and this theory was extended to bilayer graphene \cite{ando07,andokoshino09}.
An optical phonon on one graphene layer is represented by the relative displacement of two
sub-lattice atoms $A$ and $B$  as
\begin{equation}
\mathbf{u}(\mathbf{r}) = \sum_{\mathbf{q},\mu}
\sqrt{\frac{\hbar}{4NM\omega_0}}
(b_{\mathbf{q},\mu} + b^\dagger_{-\mathbf{q},\mu})
\mathbf{e}_\mu(\mathbf{q})
e^{i\mathbf{q}\cdot\mathbf{r}},
\end{equation}
where $N$ is the number of unit cells, $M$ is the mass of a
carbon atom, $\omega_0$ is the phonon frequency at the $\Gamma$ point,
$\mathbf{q} = (q_x,q_y)$ is the wave vector,
and $b_{\mathbf{q},\mu}$ and $b^\dagger_{\mathbf{q},\mu}$
are the creation and destruction operators, respectively.
The index $\mu$
represents the modes ($t$ for transverse and $l$ for longitudinal),
and corresponding unit vectors are defined by
$\vec{e}_l = i \mathbf{q}/|\mathbf{q}|$  and
$\vec{e}_t = i \hat{\mathbf{z}}\times \mathbf{q}/|\mathbf{q}|$.

The Hamiltonian of optical phonons is written as
\begin{equation}
 \mathcal{H}_{\rm ph} = \sum_{\mathbf{q},\mu} \hbar\omega_0
\left(b^\dagger_{\mathbf{q},\mu}b_{\mathbf{q},\mu} + \frac{1}{2}\right),
\end{equation}
and the interaction with an electron at the $K_+$ point is
\begin{equation}
 \mathcal{H}^K_{\rm int} =
-\sqrt{2}\frac{\beta \hbar v}{a_{\rm CC}^2}
\left[
\sigma_x u_y(\mathbf{r}) - \sigma_y u_x(\mathbf{r})
\right],
\end{equation}
where the Pauli matrix $\sigma_i$ works on the space of
$(\phi_{A1},\phi_{B1})$ for the phonon on layer 1,
and $(\phi_{A2},\phi_{B2})$ for layer 2.
The dimensionless parameter $\beta$ is related to the
dependence of the hopping integral on the interatomic distance,
and is defined by $\beta = -d \log \gamma_0 / d \log a_{\rm CC}$.
We usually expect $\beta \sim 2$.
The strength of the electron-phonon interaction
is characterized by a dimensionless parameter
\begin{equation}
 \lambda = \frac{g_vg_s}{4}\frac{36\sqrt{3}}{\pi}\frac{\hbar}{2Ma^2}
\frac{1}{\hbar\omega_0}\left(\frac{\beta}{2}\right)^2.
\end{equation}
For $M = 1.993\times 10^{23}$g and $\hbar\omega_0 = 0.196$eV
(corresponding to $1583\,$cm$^{-1}$),
we have $\lambda \approx 3\times10^{-3}(\beta/2)^2$.
For the $K_-$ point, the interaction Hamiltonian is obtained by
replacing $\sigma_i$ with $-\sigma_i^*$.

The Green's function of an optical phonon is given by a $2\times 2$ matrix
associated with phonons on layers 1 and 2. This is written as
\begin{eqnarray}
 \hat{D}(\mathbf{q},\omega) &=&
\frac{2\hbar\omega_0}
{(\hbar\omega)^2 - (\hbar\omega_0)^2
- 2\hbar\omega_0\hat{\Pi}(\mathbf{q},\omega)}
\nonumber\\
&\approx&
\frac{1}
{\hbar\omega - \hbar\omega_0 - \hat{\Pi}(\mathbf{q},\omega_0)},
\end{eqnarray}
where $\hat{\Pi}(\mathbf{q},\omega)$ is the phonon's self-energy,
and the near-equality in the second line stands because
the self-energy is much smaller than $\hbar\omega_0$.
Then, the eigenmodes are given by
eigenvectors $|u\rangle$ of ${\rm Re} \hat{\Pi}(\mathbf{q},\omega_0)$,
and the frequency shift and broadening are obtained
as the real and imaginary part of
$\langle u | \hat{\Pi}(\mathbf{q},\omega_0) |u\rangle$.

\begin{figure}
\begin{center}
\centerline{\epsfxsize=0.8\hsize \epsffile{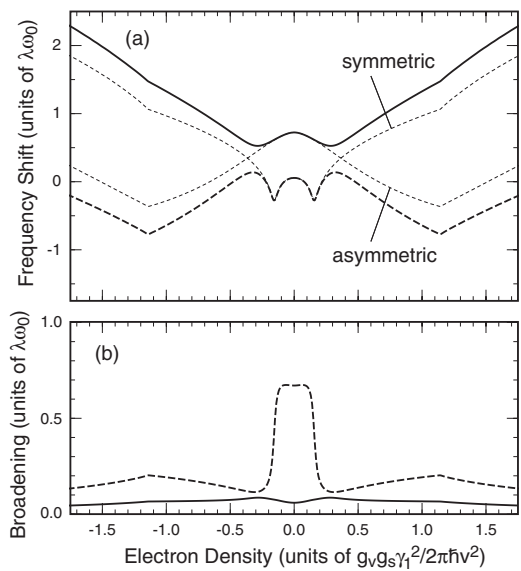}}
\caption{(a) Frequency shift and (b) broadening
of the $\Gamma$-point optical phonon in a bilayer graphene \cite{andokoshino09}.
Solid and dashed lines denote the high- and low-frequency modes,
respectively, and the thin dotted
lines in the top panel
show the frequencies for symmetric and antisymmetric modes calculated
without inclusion of their mixing.}
\label{fig_optical_phonon}
\end{center}
\end{figure}

In symmetric bilayer graphene ({\em i.e.}, no interlayer potential difference),
eigenmodes are always classified
into symmetric and antisymmetric modes
in which the displacement of the top and
bottom layers is given by $(\mathbf{u}, \mathbf{u})$
for the former, and $(\mathbf{u}, -\mathbf{u})$ for the latter.
The symmetric mode causes interband transitions between
the conduction band and valence band,
while the antisymmetric mode
contributes to the transitions between two conduction bands
(or between two valence bands).
The potential difference between two layers
gives rise to hybridisation of symmetric and antisymmetric modes \cite{ando07}.

Figure \ref{fig_optical_phonon} (a) and (b) show the calculated frequency shift and broadening,
respectively, as a function of electron density $n_s$ \cite{andokoshino09}.
Here we take $\hbar\omega_0 = \gamma_1 /2$, and introduce a phenomenological broadening factor
of $0.1 \hbar\omega_0$. We assume that $n_s$ is changed by the gate voltage,
and appropriately take account of the band deformation due to the interlayer potential difference
when $n_s \neq 0$, with the self-consistent screening effect included.
The thin dotted lines in the top panel in Figure~\ref{fig_optical_phonon} (a)
indicate the frequencies for symmetric and antisymmetric
modes calculated without inclusion of their mixing.
The lower and higher branches exactly coincide with
symmetric and asymmetric modes at $n_s=0$ where the mixing is absent.
The dip at the symmetric mode occurs when
$\hbar\omega_0 = 2\varepsilon_F$, {\em i.e.}, the interband transition
excites a valence electron exactly to the Fermi surface.
The coupling between symmetric and antisymmetric modes
arises when $n_s \neq 0$, and makes an anti-crossing
at the intersection.

\section{Electronic interactions}\label{s:eei}

Generally speaking, the low-energy behaviour of electrons
in bilayer graphene is well described by the tight-binding model
without the need to explicitly incorporate electron-electron interaction effects.
Coulomb screening and collective excitations have been described
in a number of theoretical papers \cite{wang07,sta07a,kus08,hwang08,kus09,bor-pol09,bor-pol10,sens10,abergel11,gam11}
and the importance of interaction effects in a bilayer under external gating
\cite{sta07a,castro08,dil08,sahu-min08,park10,wang-chakraborty10,cortijo-oroszlany10,toke11}
has been stressed.
Interaction effects should also be important in the presence
of a magnetic field or at very low carrier density, particularly in clean samples.

Bilayer graphene has quadratic bands which touch at low energy resulting
in a non-vanishing density of state and it
has been predicted to be unstable to electron-electron interactions at half filling.
Trigonal warping tends to cut off infrared singularities
and, thus, finite coupling strength is generally required
to realise correlated ground states;
if trigonal warping is neglected, then arbitrarily weak interactions are sufficient.
Since bilayer graphene possesses pseudospin ({\em i.e.}, which layer) and valley
degrees of freedom, in addition to real electron spin, it is possible to imagine
a number of different broken symmetry
states that could prevail depending on model details and parameter values.
Suggestions include a ferromagnetic \cite{nil06},
layer antiferromagnetic \cite{min08a,vafek10b,zhang10d,zhang11,jung11,khar11},
ferroelectric \cite{nand10a,jung11} or a charge density wave state \cite{dahal10};
topologically non-trivial phases with bulk gaps and gapless
edge excitations such as an anomalous quantum Hall state \cite{nand10b,zhang11,jung11,nand10d}
or a quantum spin Hall state \cite{zhang11,jung11} (also called a spin flux state \cite{lem12});
or a gapless nematic phase \cite{sun09,vafek10a,lem10,throck12,lem12}.

\begin{figure}[t]
\centerline{\epsfxsize=0.7\hsize \epsffile{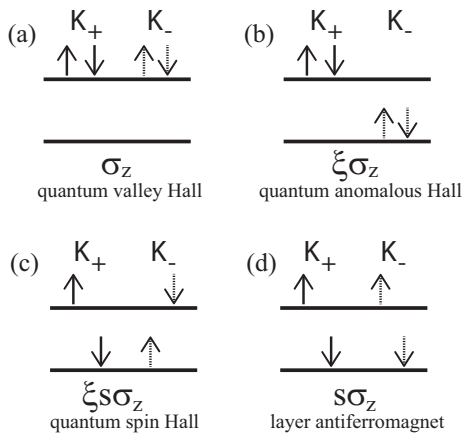}}
\caption{Schematic of the distribution of spin, valley and layer degrees of
freedom in candidates for spontaneous gapped states in bilayer
graphene \cite{min08a,zhang11,jung11}. They each contribute a term proportional
to $\sigma_z$ in the two-band Hamiltonian~(\ref{heff1}), with its
sign depending on the valley $\xi = \pm 1$ and spin $s = \pm 1$ orientation.
Arrows indicate spin orientation, located at valley $K_{+}$ (solid) or $K_{-}$ (dashed) and
on different layers (top or bottom) of a state at the top of the valence band.
}
\label{fig:states}
\end{figure}

Insulating states contribute a term proportional to $\sigma_z$ in
the two-band Hamiltonian~(\ref{heff1}) with its sign corresponding to the
distribution of layer, valley and spin degrees of freedom,
indicated in figure~\ref{fig:states}, as manifest
in their spin- and valley-dependent Hall conductivities \cite{min08a,zhang11,jung11}.
Note that the quantum spin Hall state produces a term equivalent to that
of intrinsic spin-orbit coupling, equation~(\ref{hso}), which
describes a topological insulator state \cite{kanemele05,cort10,prada11}.
By way of contrast, the gapless nematic phase has a qualitatively similar effect on the spectrum of
bilayer graphene as lateral strain \cite{mucha11}, described in
section~\ref{s:pas}, producing an additional
term in the two-band Hamiltonian of the form of equation~(\ref{hstrain}),
with parameter $w$ taking the role of an order parameter.

Experiments on suspended bilayer graphene devices have found
evidence for correlated states at very low density and zero magnetic field \cite{weitz10,martin10,mayorov11,freitag12,velasco12,bao12,vel12}.
Conductivity measurements of double-gate devices \cite{weitz10}
observed a non-monotonic dependence of the resistance on electric field
cumulating in a non-divergent resistance at zero field
while compressibility measurements of single-gate devices \cite{martin10}
found an incompressible region near the charge neutrality point.
These measurements were interpreted as being
consistent with either the anomalous quantum Hall state,
in which the separate layers of the device are valley polarised,
or the gapless nematic phase.

Subsequently, conductivity measurements on single-gate devices \cite{mayorov11}
observed a weak temperature dependence of the width of the conductivity
minimum near zero carrier density, suggesting a suppressed
density of states as compared to that expected for a parabolic
dispersion, as well as particularly robust cyclotron gaps at filling factor $\nu = \pm 4$,
observations attributed to the presence of the nematic phase.

However, other experiments \cite{freitag12,velasco12,bao12,vel12} observed insulating
states indicating the formation of ordered phases with energy gaps of about $2\,$meV.
There was evidence for the existence of edge states in one of these
experiment \cite{freitag12}, but not in the others \cite{velasco12,bao12,vel12},
and the latter observations including their response to perpendicular electric field \cite{velasco12,bao12} and tilted magnetic field \cite{vel12} were seen as being consistent with
a layer antiferromagnetic state, in which the separate layers of the device are
spin polarised, figure~\ref{fig:states}(d).
At present, there appear to be contradicting experimental and theoretical results, but it should be
noted that renormalisation group studies \cite{throck12,lem12,zhang-min12}
have highlighted the sensitivity of the phase diagram of bilayer graphene to microscopic details.

In the absence of interactions, the low-energy Landau level spectrum of bilayer
graphene \cite{mcc06a} consists of a series of fourfold
(spin and valley) degenerate levels with an eightfold-degenerate level
at zero energy, as described in section~\ref{s:qhe}. The resulting Hall conductivity
consists of a series of plateaus at conventional integer positions
of $4 e^2/h$, but with a double-sized step of $8 e^2/h$
across zero density \cite{novo06}.
Interaction effects are expected to lift the
level degeneracy of quantum Hall ferromagnet states at
integer filling factors \cite{nomura06,barlas08,shi09,abanin09,gor10,gor11}.
Indeed, an insulating state at filling factor $\nu = 0$
and complete splitting of the eightfold-degenerate level at zero energy
have been observed with quantum states at filling factors
$\nu = 0, ±1, ±2$ and $±3$ in high-mobility suspended bilayer graphene
at low fields (with all states resolved at $B = 3\,$T) \cite{Feld09a,vanelf12}
and in samples on silicon substrates at high fields,
typically above $20\,$T \cite{zhao10,kim11}.

The fractional quantum Hall effect has been observed recently in
monolayer graphene, both in suspended samples \cite{du09,bol09}
and graphene on boron nitride \cite{dean11}, and there is evidence for
it in bilayer graphene \cite{bao10}, too.
Strongly-correlated states at fractional filling factors and the prospect of tuning
their properties has been the focus of recent
theoretical attention \cite{shib09,apalkov10,apalkov11,pap11,sniz11}.
Clearly, the nature of the electronic properties
of bilayer graphene in high-mobility samples is a complicated problem,
and it is likely to be an area of further intense experimental and
theoretical investigation in the following years.

\section{Summary}

This review focused on the single-particle theory of
electrons in bilayer graphene, in the shape of the tight-binding
model and the related low-energy effective Hamiltonian.
Bilayer graphene has two unique properties:
massive chiral quasiparticles in two parabolic
bands which touch at zero energy,
and the possibility to control an infrared gap between these
low-energy bands by applying an external gate potential.
These features have a dramatic impact on many physical properties
of bilayer graphene including some described here: optical and transport properties,
orbital magnetism, phonons and strain.
A number of topics were not covered here in great detail or at all; we refer
the reader to relevant detailed reviews of graphene including
electronic transport \cite{peres10,dassarma11},
electronic and photonic devices \cite{avouris10},
scanning tunnelling microscopy \cite{andrei12},
Raman spectroscopy \cite{ferrari-review,mal-pim09},
magnetism \cite{yaz10}, spintronics and pseudospintronics \cite{pes12},
Andreev reflection at the interface with a superconductor and Klein tunnelling \cite{been08},
growth and applications \cite{novo12},
and the properties of graphene in general \cite{cnreview,abergel10e}.
Finally, although the central features of the single-particle theory are already
established,
the same can not be said of the influence of electronic interactions,
which is likely to remain a subject of intense research,
both theoretical and experimental, in the near future at least.

The authors gratefully acknowledge colleagues for fruitful collaboration in graphene
research, in particular T. Ando and V. I. Fal'ko, and we acknowledge
funding by the JST-EPSRC Japan-UK Cooperative Programme Grant EP/H025804/1.

\end{document}